\documentclass[manuscript]{aastex61}

\usepackage{natbib}

\newcommand{\kms}{km\,s$^{-1}$}
\newcommand{\arcpx}{$''$ pixel$^{-1}$}

\received{July 1, 2017}
\revised{December 27, 2017}
\accepted{\today}
\submitjournal{AAS Journals}

\shorttitle{SN 2010bt}
\shortauthors{Elias-Rosa et al.}

\begin{document}

\title{The Type IIn Supernova 2010bt: The Explosion of a Star in Outburst}

\correspondingauthor{Nancy Elias-Rosa}
\email{eliasrosa.astro@gmail.com}

\author{Nancy Elias-Rosa}
\affiliation{INAF - Osservatorio Astronomico di Padova, vicolo dell'Osservatorio 5, Padova I-35122, Italy}
\affiliation{Institut de Ci\`encies de l'Espai (CSIC-IEEC), Campus UAB, Cam\'{\i} de Can Magrans S/N, 08193 Cerdanyola (Barcelona), Spain}

\author{Schuyler D.~Van Dyk}
\affiliation{Caltech/IPAC, Mailcode 100-22, Pasadena, CA 91125, USA.}

\author{Stefano Benetti}
\affiliation{INAF - Osservatorio Astronomico di Padova, vicolo dell'Osservatorio 5, Padova I-35122, Italy}

\author{Enrico Cappellaro}
\affiliation{INAF - Osservatorio Astronomico di Padova, vicolo dell'Osservatorio 5, Padova I-35122, Italy}

\author{Nathan Smith}
\affiliation{Steward Observatory, University of Arizona, Tucson, AZ 85720, USA.}

\author{Rubina Kotak}
\affiliation{Astrophysics Research Centre, School of Mathematics and Physics, Queen's University Belfast, Belfast BT7 1NN, UK.}

\author{Massimo Turatto}
\affiliation{INAF - Osservatorio Astronomico di Padova, vicolo dell'Osservatorio 5, Padova I-35122, Italy}

\author{Alexei V.~Filippenko}
\affiliation{Department of Astronomy, University of California, Berkeley, CA 94720-3411, USA.}
\affiliation{Miller Senior Fellow, Miller Institute for Basic Research in Science, University of California, Berkeley, CA 94720, USA.}

\author{Giuliano Pignata}
\affiliation{Departamento de Ciencias F\'{i}sicas, Universidad Andres Bello, Avda. Rep\'ublica 252, Santiago, 8320000, Chile.}
\affiliation{Millennium Institute of Astrophysics (MAS), Nuncio Monse\~nor S\'otero Sanz 100, Providencia, Santiago, Chile.}

\author{Ori~D.~Fox}
\affiliation{Space Telescope Science Institute, 3700 San Martin Drive, Baltimore, MD 21218, USA.}

\author{Lluis Galbany}
\affiliation{PITT PACC, Department of Physics and Astronomy, University of Pittsburgh, Pittsburgh, PA 15260, USA.}

\author{Santiago Gonz\'alez-Gait\'an}
\affiliation{CENTRA, Departamento de F\'isica, Instituto Superior T\'ecnico, Universidade de Lisboa, Avenida Rovisco Pais 1, 1049 Lisboa, Portugal.}

\author{Matteo~Miluzio}
\affiliation{European Space Astronomy Centre (ESAC), European Space Agency, Villanueva de la Ca\~nada, Madrid, E-28692, Spain}

\author{L.~A.~G.~Monard}
\affiliation{Kleinkaroo Observatory, Center for Backyard Astronomy Kleinkaroo, Sint Helena 1B, PO Box 281, Calitzdorp 6660, South Africa.}

\author{Mattias Ergon}
\affiliation{The Oskar Klein Centre, Department of Astronomy, AlbaNova, Stockholm University, 10691 Stockholm, Sweden.}

\begin{abstract}

It is well known that massive stars ($M > 8$ M$_{\sun}$) evolve up to the collapse of the stellar core, resulting in most cases as a supernova (SN) explosion. Their heterogeneity is related mainly to different configurations of the progenitor star at the moment of the explosion, and to their immediate environments. We present photometry and spectroscopy of SN~2010bt, which was classified as a Type~IIn~SN from a spectrum obtained soon after discovery and was observed extensively for about two months. After the seasonal interruption owing to its proximity to the Sun, the SN was below the detection threshold, indicative of a rapid luminosity decline. We can identify the likely progenitor with a very luminous star (log $L/{\rm L}_{\sun} \approx 7$) through comparison of {\sl Hubble Space Telescope\/} images of the host galaxy prior to explosion with those of the SN obtained after maximum light. Such a luminosity is not expected for a quiescent star, but rather for a massive star in an active phase. This progenitor candidate was later confirmed via images taken in 2015 ($\sim 5$~yr post-discovery), in which no bright point source was detected at the SN position. 
Given these results and the SN behavior, we conclude that SN~2010bt was likely a Type IIn SN and that its progenitor was a massive star that experienced an outburst shortly before the final explosion, leading to a dense H-rich circumstellar environment around the SN progenitor.

\end{abstract}


\section{Introduction}\label{intro}

Type II ``narrow'' (IIn) supernovae (SNe) are a heterogeneous subset of core-collapse (CC) SNe \citep{schlegel90}. According to \citet{li11,smith11c}, they represent 9\% of all CC-SNe and seem to preferentially occur in small and late-type spiral galaxies.

This subclass of objects can be distinguished from other types of CC-SNe by their spectral appearance. On the other hand, they can also be confused with nonterminal outbursts of very massive stars. The broad absorption lines typical of many SNe are weak or absent in SNe~IIn throughout their evolution. Instead, they show strong, narrow Balmer emission components (with full width at half-maximum intensity [FWHM] ranging from a few tens to a few hundreds of \kms) atop broader emission (which can have intermediate-velocity components with FWHM $\approx 1000$ \kms,  as well as  broad components with FWHM of a few thousands of \kms; see, e.g., \citealt{filippenko97}). The narrow lines are thought to arise from the surrounding circumstellar material (CSM) ionized by the shock-interaction emission (e.g., \citealt{chugai94}). Their light curves, however, exhibit a wide range of properties (see, e.g., \citealt{smith17}). This diversity is related to the mass loss history during the evolution of massive stars. 

It was suggested that the progenitors of a fraction of these interacting SNe are massive stars in a luminous blue variable (LBV) phase. These stars are among the most luminous ($M_{\rm bol} < -9.6$ mag) and massive ($> 50$ M$_{\sun}$) stars in late-type galaxies (e.g., \citealt{humphreys94}). The evidence arises from the identification of the progenitor stars of few SNe~IIn, such as SN~2005gl \citep{galyam07,galyam09}, SN~2009ip \citep{smith10,foley11,smith14}, and SN~2015bh \citep{eliasrosa16,thone17}. A luminous, blue point-like source, originating from either a young cluster or a single star, was first proposed as the progenitor of SN~2010jl by \citet{smith11}. Subsequently, \citet{fox17} argued that this source was most likely a massive young cluster, although they did not rule out the possibility of a very luminous progenitor star obscured by dust (see also \citealt{dwek17}). However, the nature of these progenitors is not fully clear, and other precursor channels have been proposed (see, e.g., \citealt{mauerhan12} in the case of the SN~1998S, or \citealt{smith14c} for a larger review of this topic).

As mentioned above, the powerful nonterminal outbursts of very massive stars can mimic genuine SN explosions, in terms of energetics and spectral appearance (both show incipient narrow lines of hydrogen in emission); thus, they are usually referred to as ``SN impostors'' (e.g., SN~1997bs, \citealt{vandyk00}; SN~2000ch, \citealt{pastorello10}; see also the general discussions in \citealt{smith11b} and \citealt{vandyk12}). Discriminating between SN impostors and SNe~IIn is challenging, and for several objects both possibilities remain viable (see, e.g., the case of SN~2007sv; \citealt{tartaglia15}).

\begin{figure}
\figurenum{1}
\plotone{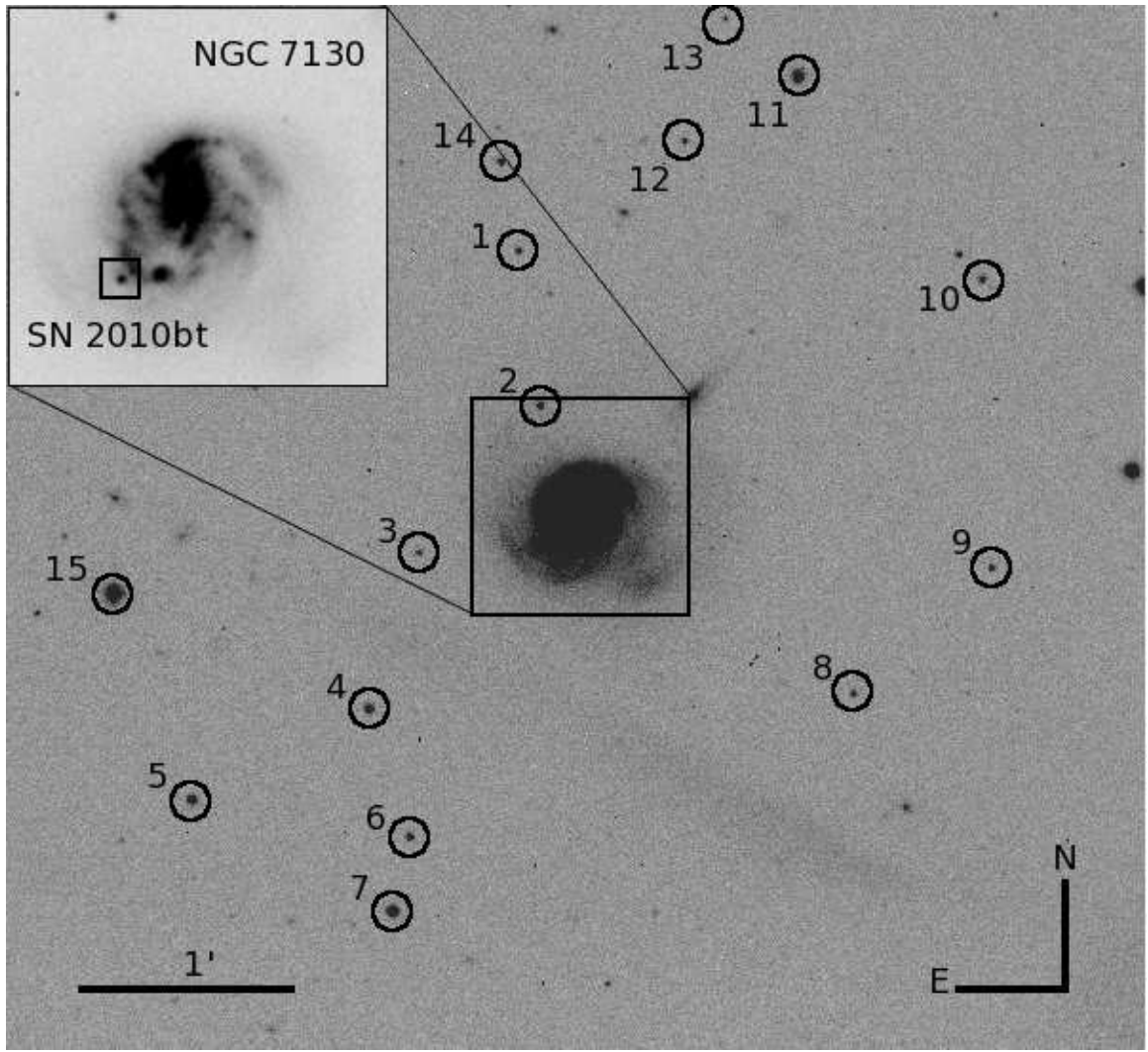}
\epsscale{0.4}
\caption{$V$-band image of SN~2010bt in NGC~7130 obtained with the 1.3~m SMARTS telescope+ANDICAM at CTIO on 2010 May 18 (field of view $\sim 6\arcmin \times 6\arcmin$). The SN and local photometric sequence stars are indicated.
\label{fig_seq}}
\end{figure}

SN~2010bt was discovered in NGC~7130 on 2010 April 17.11 (UT dates are used throughout this paper) at an unfiltered magnitude of 15.9, and confirmed on 2010 April 18.14 at a magnitude of 15.8 (\citealt{monard10}; Figure \ref{fig_seq}). The presence of Balmer emission with multicomponent profiles in the spectrum from 2010 April 18.39 led to the classification of SN~2010bt as a SN~IIn, a few days after discovery \citep{turatto10}. 

NGC~7130 (or IC~5135; $z = 0.016$\footnote{NED, NASA/IPAC Extragalactic Database; 
\url{http://nedwww.ipac.caltech.edu/}.}; morphological type Sa pec) was classified as a Seyfert 2 galaxy by 
\citet{phillips83} --- i.e., an active galactic nucleus (AGN) with obscuring material that precludes a direct view of its nuclear region. Throughout the paper, we adopt a distance to NGC~7130 of 65.4 $\pm$ 4.6 Mpc ($\mu = 34.08 \pm 0.15$ mag), resulting from the recession velocity of the galaxy \citep{deVauco76} corrected for Local Group infall into the Virgo cluster \citep{mould00}, $v_{\rm Vir} = 4771 \pm 17$ \kms\ ($z = 0.016$), assuming H$_0 = 73$ \kms\,Mpc$^{-1}$ (values taken from NED).

This paper presents and discusses our photometric and spectroscopic observations of SN~2010bt in Section \ref{sn10bt}. In Section \ref{identification}, we compare pixel-by-pixel four sets of {\sl Hubble Space Telescope\/} ({\sl HST}) data, and we discuss the nature of the progenitor star in Section \ref{natureprog}. We conclude in Section \ref{conclusions}.

%
\section{The Nature of SN~2010bt}\label{sn10bt}

\subsection{Photometry}\label{SNph}

Optical $BVRI$ images of SN~2010bt were obtained with the 1.3~m Small \& Moderate Aperture Research Telescope System (SMARTS)+ANDICAM at Cerro Tololo Inter-American Observatory (CTIO), the 6.5~m Magellan Clay Telescope+LDSS-3 at Las Campanas Observatory, and the 3.58~m New Technology Telescope (NTT)+EFOSC2 at the European Southern Observatory (ESO) of La Silla Observatory, all located in Chile. The data were obtained thanks to a collaboration between American and European researchers. We also include in our dataset unfiltered data from the Bronberg Observatory (South Africa) and archival images from the 1.0~m Jacobus Kapteyn Telescope at Roque de Los Muchachos Observatory (Spain). More information about the telescopes and instruments used in this follow-up campaign can be found in Table \ref{table_setup}.

The photometric observations were processed following the standard recipe in {\sc iraf} for CCD images (trimming, overscan, bias, and flat-field corrections). Due to the location of SN~2010bt in NGC~7130, contamination of the SN photometry by the host-galaxy light was a serious problem. We therefore used the template-subtraction technique to remove this background and hence to measure more accurately the SN magnitudes. The template images of NGC~7130 were obtained with the 1.3~m SMARTS telescope+ANDICAM at CTIO on 2011 August 03. Each SN image was registered geometrically and photometrically with their corresponding template using a dedicated pipeline (SNOoPY). This consists of a collection of {\sc python} scripts calling standard {\sc iraf} tasks (through {\sc pyraf}), and other specific analysis tools, in particular {\sc sextractor}, for source extraction and star/galaxy separation. The instrumental magnitudes of the SN and the reference stars in the SN field were measured in the subtracted images (produced with {\sc hotpants}) using the point-spread-function (PSF) fitting technique with the {\sc daophot} package. In order to calibrate the instrumental magnitudes to the standard photometric \textsc{vegamag} system, we used the magnitudes and colors of 15 local sequence stars in the SN field (Figure \ref{fig_seq} and Table \ref{table_seq}). The unfiltered data were transformed to the Johnson-Cousins $R$ band, for which the effective wavelength is similar to the natural instrumental band defined by the CCD quantum efficiency of the detector that was used.

Near-infrared (NIR) observations were also obtained with the 3.6~m NTT+SOFI at the ESO Observatory of La Silla. We include in this work the data obtained at the 8.2~m unit telescope UT4 of the Very Large Telescope + HAWK-I at the ESO Observatory of Cerro Paranal (PI E. Cappellaro, 083.D-0259(A)), presented by \citet{miluzio13}. The data were reduced using standard procedures for the VLT+HAWK-I data (see \citealt{miluzio13} for details on the data reduction). For the NIR images we used two set of images of the field as templates to subtract from the SN images, depending on the instrumentation, e.g., images from ESO NTT+SOFI taken on 2004 October 04 (PI P. Lira, 074.B-0375(A)) and images from ESO VLT UT4+HAWK-I on 2011 April 24 (PI E. Cappellaro, 083.D-0259(A)). The instrumental $JHK_s$ photometry was calibrated using Two Micron All Sky Survey (2MASS\footnote{\url{http://www.ipac.caltech.edu/2mass/}.}) stars in the field.

When the transient was not detected, upper limits were adopted corresponding to 2.5 times the standard deviation in the background. Uncertainty estimates were obtained through an artificial star experiment, combined (in quadrature) with the PSF fit error returned by {\sc daophot} and the propagated uncertainties from the photometric calibration.

The final {\sc vegamag} calibrated magnitudes of SN~2010bt are listed in Tables \ref{table_JCph} and \ref{table_NIRph}. The single epoch with magnitudes in the Sloan system were transformed to the {\sc vegamag} system and included in Table \ref{table_JCph} by employing the relations given by \citet{blanton07}.

Space-based optical and ultraviolet (UV) data taken with the Ultraviolet/Optical Telescope (UVOT) onboard the Neil Gehrels Swift Observatory ({\sl Swift\/}) complement our ground-based photometry. The calibrated SN images in the UVOT-Vega system were obtained from SOUSA ({\sl Swift}'s Optical/Ultraviolet Supernova Archive; \citealt{brown14}). Upper limits corresponding to 3 times the standard deviation in the background were estimated when the transient was not detected. The {\sl Swift\/} photometry for SN~2010bt is reported in Table \ref{table_UVph}.

Finally, {\sl HST} observed the SN~2010bt field at four epochs: 1994, 2003, 2010, and 2015 (see Sections \ref{identification} and \ref{natureprog} for more details). The \textsc{vegamag} magnitudes of the progenitor candidate and transient were obtained using Dolphot (see Table \ref{table_progmag}). 
\\

\begin{figure}
\figurenum{2}
\plotone{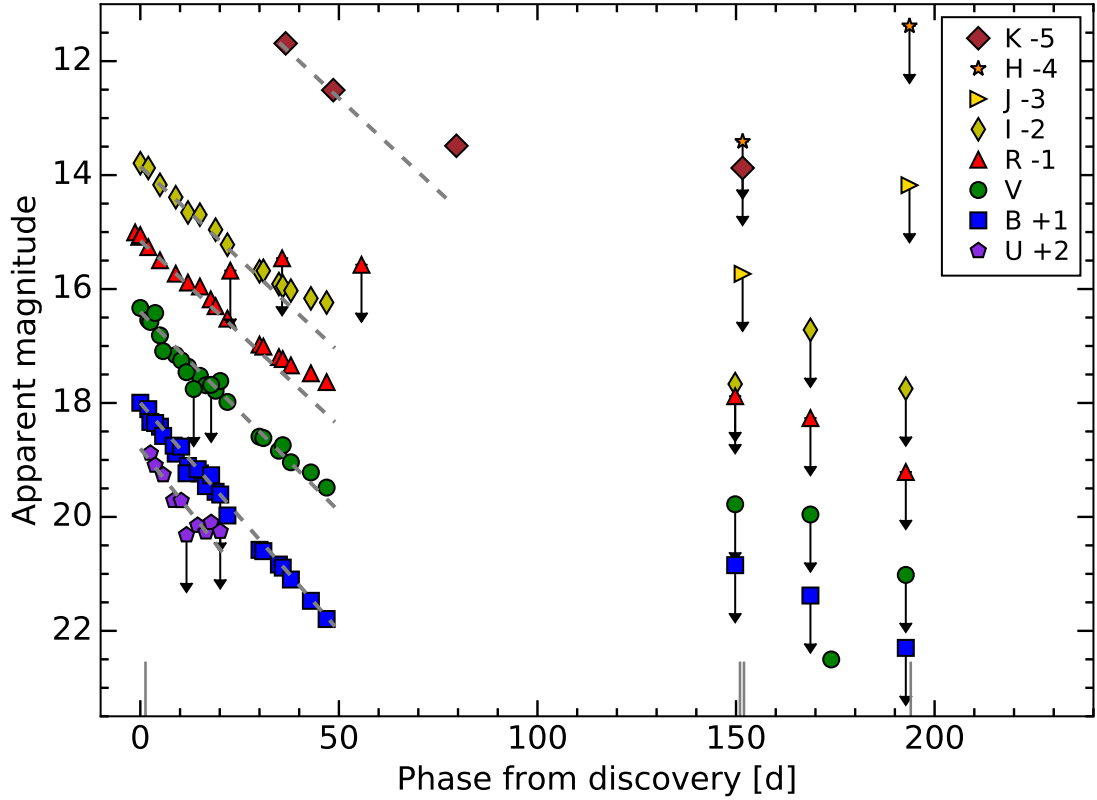}
\caption{Optical and NIR light curves of SN~2010bt. Upper limits are indicated by symbols with arrows. The solid marks on the abscissa indicate the phases at which spectra were obtained. The dashed lines show the slopes of the light curves during the first 20 days from discovery. The light curves have been shifted for clarity by the amounts indicated in the legend. The uncertainties for most of the data points are smaller than the plotted symbols. 
\label{fig_ph}}
\end{figure}

SN~2010bt was observed during the first $\sim 50$ (optical) to 80 (NIR) d after discovery. Subsequently it came too close to the Sun's direction and was lost. The telescopes could point to the field again around two months after the last observation in the NIR; however, at that time the SN was no longer visible (at least with ground-based telescopes). Instead, it was detected in a F606W ($\sim V$)-band {\sl HST} exposure obtained $\sim 175$ d after discovery. The $UBVRIJHK$ light curves, including the $UBV$-UVOT data, are shown in Figure~\ref{fig_ph}. In the figure, phase is relative to the discovery date on 2010 April 17.10, or MJD 55303.1. However, in the following, and in order to better compare SN~2010bt with other SNe, we will refer to the $V$ maximum computed as the discovery date minus 10 d (see Section \ref{SNexpl} for more details). 

As one can see in Figure~\ref{fig_ph}, SN~2010bt was discovered after maximum light. The light curves show a decline similar to those of the rapidly declining SNe~II studied by \citealt{anderson14} (with an average $V$-band light-curve decline $> 1.3$ mag in the first 50 d after peak). In fact, the $UBV$ light curves of SN~2010bt show constant decline rates of 4.3, 4.0, and 3.3 mag/50~d\footnote{Considering the interval from $\sim 1$ d to 50 d after the discovery date.}, respectively, while the redder light curves ($RI$) slightly flatten out in brightness around day 30 after SN discovery. The $K$-band light curve was also slowly declining (2.0 mag/50~d, considering the interval from $\sim 38$ d to 80 d after the discovery date). SN~2010bt was only detected in two epochs in the {\sl Swift} $UVW1$ band, with a similar decline to that of the bluer light curves ($\sim 3.4$ mag/50~d). 

\begin{figure}
\figurenum{3}
\plotone{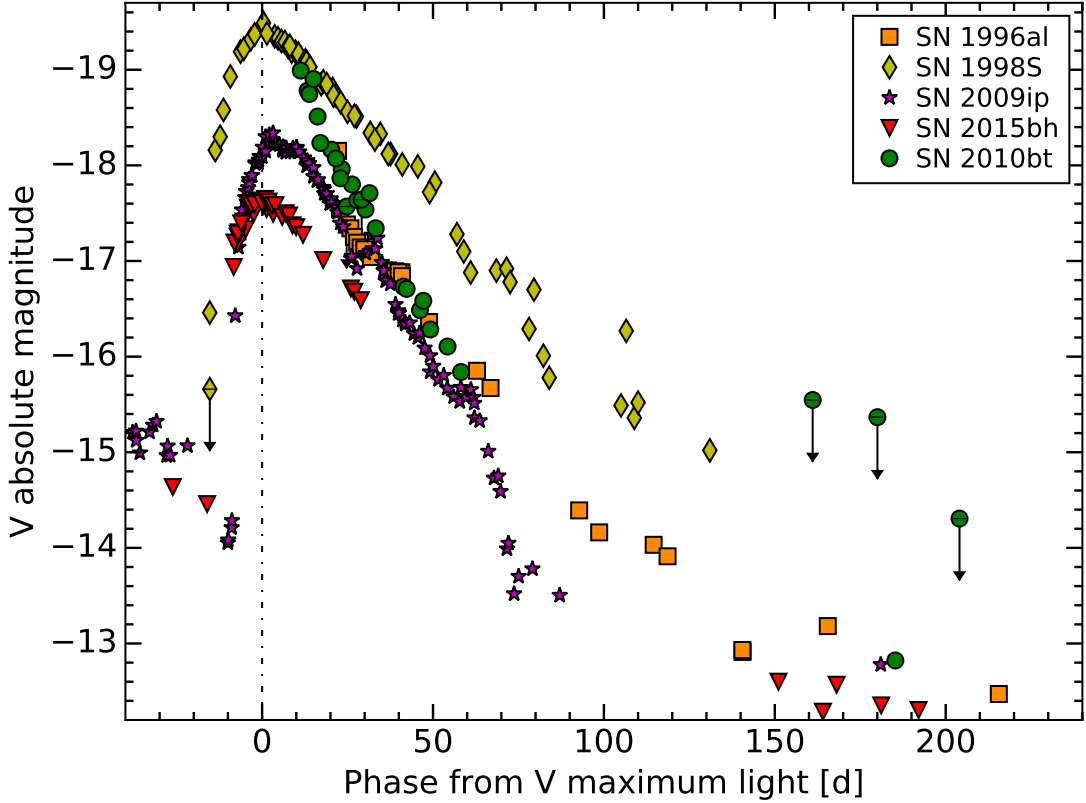}
\caption{Absolute $V$ light curve of SN~2010bt ({\it circles}) along with those of the SN~IIn~1996al ({\it squares}), SN~IIn~1998S ({\it diamonds}), the controversial transient SN~2009ip ({\it 5-pointed stars}), and SN~2015bh ({\it inverted triangles}). Upper limits are indicated by symbols with arrows. The dot-dashed vertical line indicates the $V$-band maximum light of SN~2010bt. Ages are relative to $V$ maximum light (we have assumed $V$ maximum date = discovery date $- 10$ d, for SN~2010bt). 
\label{fig_abs}}
\end{figure}

Figure \ref{fig_abs} shows a comparison between the evolution of the absolute $V$ magnitude of SN~2010bt, compared with the SNe~IIn~1996al \citep{benetti16}, 1998S \citep{liu00,fassia00,pozzo04}, and the SNe~IIn/SN impostors 2009ip \citep{pastorello13,fraser13,mauerhan13a,margutti14} and 2015bh \citep{eliasrosa16}. The comparison SNe have been corrected for extinction using published estimates and assuming the \citet{cardelli89} extinction law (see also Table \ref{tabla_SNe}). SN~2010bt exhibits a rapid decline at early times, similar to the decline from the ``b'' event of SN~2009ip and slightly faster than SN~1996al. This decline rate is also reminiscent of rapidly declining SNe~II or even of SNe~IIb (see, e.g., \citealt{li11,anderson14}). 
The absolute $V$ magnitude at maximum of SN~2010bt was $-19$ mag or brighter, consistent with both the typical $V$-band peak magnitudes of SNe~IIn ($M_V = -18.4 \pm 1.0$ mag; \citealt{kiewe12}) and of SNe~II-P/L ($M_V = -16.89 \pm 0.98$ mag; \citealt{galbany16}).

\begin{figure}
\figurenum{4}
\epsscale{0.8}
\plotone{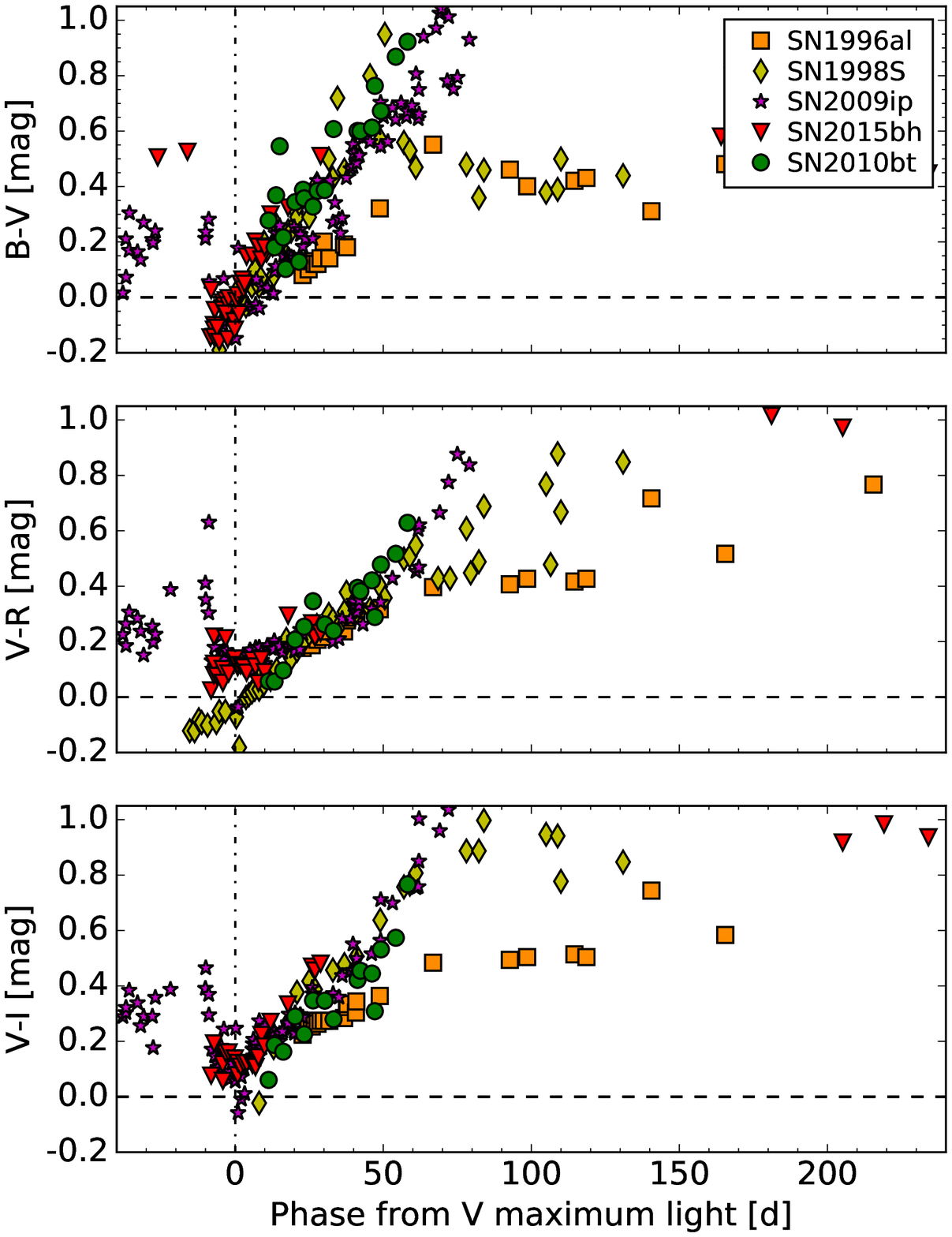}
\caption{Intrinsic color curves of SN~2010bt ({\it circles}), compared with those of SNe~1996al ({\it squares}), 1998S ({\it diamonds}), 2009ip ({\it 5-pointed stars}), and 2015bh ({\it inverted triangles}). The dot-dashed vertical line indicates the $V$-band maximum light of SN~2010bt. Ages are relative to $V$ maximum light (we have assumed $V$ maximum date = discovery date $- 10$ d, for SN~2010bt). 
\label{fig_color}}
\end{figure}

The early-time $(B-V)_0$, $(V-R)_0$, and $(R-I)_0$ color curves of SN~2010bt (see Figure \ref{fig_color}) exhibit rapid evolution from blue to red, similar to the color evolution of the other SNe~IIn, yet different from SN~1996al. 

We have computed the pseudobolometric light curve of SN~2010bt by integrating the flux at different wavelengths derived from the extinction-corrected optical apparent magnitudes over the sparse observations in the $UVW1$ through $K$ bands. Fluxes were measured considering only the epochs when $V$-band observations were available. When photometric measurements in one band at given epochs were not available, the flux was estimated by interpolating magnitudes from epochs close in time or, when necessary, by extrapolating the missing photometry assuming a constant color. We estimated the pseudobolometric flux at each epoch by integrating the spectral energy distribution (SED) using the trapezoidal rule and assuming zero flux outside the integration boundaries. Finally, the effective fluxes were converted to luminosities using the adopted distance to the SN (see Section~\ref{intro}). The errors in the bolometric luminosity include the uncertainties in the distance estimate, the extinction, and the apparent magnitudes. Note that the $UVW1$ band provides about 13\% of the luminosity.

\begin{figure}
\figurenum{5}
\plotone{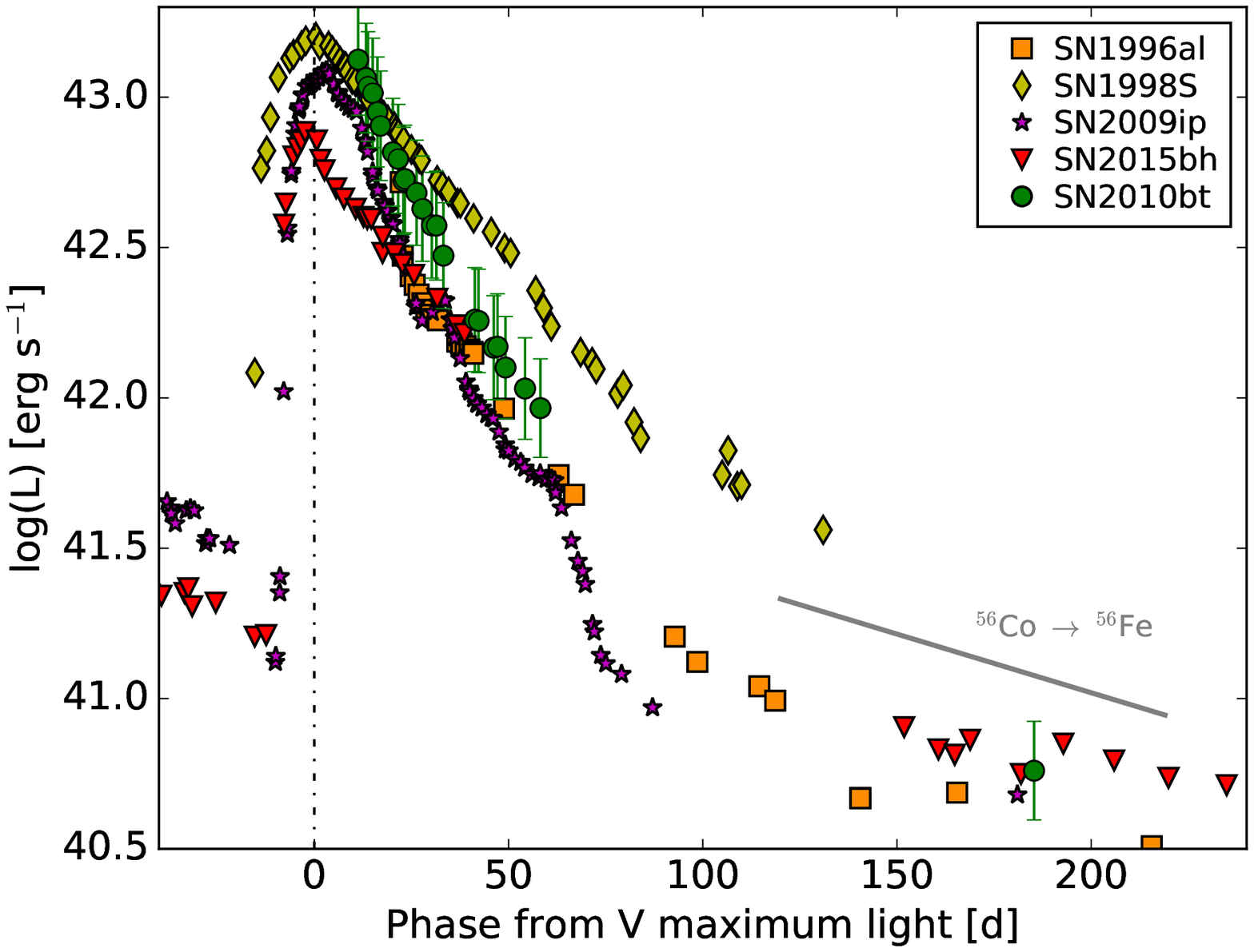}
\caption{Pseudobolometric light curves of SN~2010bt ({\it circles}), compared with those of SNe~1996al ({\it squares}), 1998S ({\it diamonds}), 2009ip ({\it 5-pointed stars}), and 2015bh ({\it inverted triangles}). The dot-dashed vertical line indicates the $V$-band maximum light of SN~2010bt. Ages are relative to $V$ maximum light (we have assumed $V$ maximum date = discovery date $- 10$ d, for SN~2010bt). 
\label{fig_bol}}
\end{figure}

In Figure \ref{fig_bol} we present the pseudobolometric light curve of SN~2010bt as well as those of SNe~1996al, 1998S, 2009ip, and 2015bh computed with the same prescriptions (for SNe~1996al and 1998S we have no observations in UV bands). As shown earlier, the luminosity decline of SN~2010bt is similar to that of SN~2009ip, while it was more luminous than all other SNe in the sample, with the possible exception of SN~1998S. Considering the first epoch in the $V$ band for SN~2010bt, its peak may have reached a luminosity $> 1.3 \times 10^{43}$ erg s$^{-1}$ (SN~1998S had a luminosity at peak of $1.6 \times 10^{43}$ erg s$^{-1}$). 

Assuming that the tail of SN~2010bt followed the radioactive $^{56}$Co decay with full trapping, and considering the explosion date to be between 15 and 50 d before the discovery date (see Section \ref{SNexpl}) and the {\sl HST\/} detection in F606W ($\sim V$) at 175.3 d (from discovery), we can roughly estimate $\leq 0.03 \pm 0.01$ M$_{\sun}$ for the $^{56}$Ni mass, using the formula given by \citet{hamuy03}. This value is at the low end in comparison with typical values for CC-SNe, 0.001--0.3 M$_{\sun}$ \citep{hamuy03}. The $^{56}$Ni mass of SN~2010bt is also in agreement to what was estimated for the interacting SNe~1996al, 2009ip, and 2015bh ($M_{\rm Ni} \le 0.02$, $\le 0.08$, and $\le 0.04$ M$_{\odot}$, respectively; \citealt{benetti16,fraser13,margutti14,smith14,eliasrosa16}).

\begin{deluxetable}{@{}lllll@{}}
\rotate
\tabletypesize{\footnotesize} 
\tablenum{1}
\tablewidth{0pt}
\tablecolumns{5}
\tablecaption{Basic Information about Telescopes and Instruments Used
\label{table_setup}}
\setlength{\tabcolsep}{0.03in}
\tablehead{
\colhead{Table key\tablenotemark{a}} & \colhead{Telescope} & \colhead{Instrument} & \colhead{Pixel Scale} & \colhead{Location} \\
\colhead{} & \colhead{} & \colhead{} & \colhead{($''$ pixel$^{-1}$)} & \colhead{}}
\startdata
ANDICAM & 1.3~m SMARTS\tablenotemark{b} & ANDICAM & 0.37 & CTIO\tablenotemark{c}, Chile \\ 
BO & 0.3~m Meade RCX400 f/8 telescope & SBIG ST8-XME CCD & 1.37 & Bronberg Observatory, South Africa \\
EFOSC2 & 3.6~m New Technology Telescope & EFOSC2 & 0.12 & ESO\tablenotemark{d}, La Silla Obs., Chile \\
HAWK-I & 8.2~m Very Large Telescope-UT4 & HAWK-I & 0.11 & ESO, La Silla Obs., Chile \\
HST\_WFPC2 & 2.4~m {\sl HST} & WFPC2 & 0.05\tablenotemark{e} & \nodata \\
HST\_ACS/HRC & 2.4~m {\sl HST}  & ACS/HRC & 0.03 & \nodata \\
HST\_ACS/WFC & 2.4~m {\sl HST}  & ACS/WFC & 0.05 & \nodata \\
HST\_WFC3 & 2.4~m {\sl HST} & WFC3/UVIS & 0.04 & \nodata \\
JAG & 1.0~m Jacobus Kapteyn Telescope & JAG\tablenotemark{f} & 0.33 & ORM\tablenotemark{g}, La Palma, Spain\\
LDSS-3 & 6.5~m Magellan Clay Telescope & LDSS-3 & 0.19 & Las Campanas Obs., Chile\\
SOFI & 3.6~m New Technology Telescope & SOFI & 0.29  & ESO, La Silla Obs., Chile \\
Spitzer & 0.8~m {\sl Spitzer Space Telescope} & IRAC & 0.60 & \nodata \\
Swift & 0.3~m Ritchey-Chretien UV/optical Telesc. & {\sl Swift} & 0.50 & \nodata \\
\enddata
\tablenotetext{a}{See Tables~\ref{table_JCph}, \ref{table_NIRph}, and \ref{table_UVph}.}
\tablenotetext{b}{Small \& Moderate Aperture Research Telescope System.}
\tablenotetext{c}{Cerro Tololo Inter-American Observatory.}
\tablenotetext{d}{European Southern Observatory.}
\tablenotetext{e}{WFPC2 contains four chips. The SN~2010bt field was observed with the Planetary Camera ($0.05''$ pixel$^{-1}$).}
\tablenotetext{f}{JKT Acquisition and Guidance Unit.}
\tablenotetext{g}{Observatorio del Roque de los Muchachos.}
\end{deluxetable}

\begin{deluxetable}{@{}lccccccc@{}}
\tablenum{2}
\tablewidth{0pt}
\tablecolumns{8}
\tablecaption{Magnitudes of the Local Comparison Stars\tablenotemark{a}
\label{table_seq}}
\tablehead{
\colhead{Star} & \colhead{$B$} & \colhead{$V$} & \colhead{$R$} & \colhead{$I$} & \colhead{$J$} & \colhead{$H$} & \colhead{$K$}\\
\colhead{} & \colhead{(mag)} & \colhead{(mag)} & \colhead{(mag)} & \colhead{(mag)} & \colhead{(mag)} & \colhead{(mag)} & \colhead{(mag)}}
\startdata
 1 & 19.86 (0.07) & 19.04 (0.05) & 18.51 (0.03) & 17.96 (0.04) & \nodata & \nodata & \nodata \\
 2 & 20.24 (0.07) & 18.70 (0.05) & 17.60 (0.07) & 16.19 (0.04) & 14.81 (0.03) & 14.14 (0.02) & 13.85 (0.04) \\
 3 & 21.28 (0.07) & 19.66 (0.04) & 18.30 (0.04) & 16.51 (0.05) & 14.84 (0.03) & 14.20 (0.04) & 13.92 (0.05) \\
 4 & 18.68 (0.04) & 17.80 (0.04) & 17.20 (0.03) & 16.62 (0.03) & 15.91 (0.07) & 15.40 (0.09) & 15.38 (0.13) \\
 5 & 18.13 (0.04) & 17.40 (0.03) & 16.92 (0.02) & 16.51 (0.03) & 15.90 (0.07) & 15.54 (0.08) & 15.43 (0.14) \\
 6 & 19.23 (0.04) & 18.28 (0.04) & 17.65 (0.04) & 17.10 (0.03) & 16.56 (0.11) & 15.96 (0.11) & 15.52 (0.17) \\
 7 & 17.31 (0.03) & 16.49 (0.03) & 16.00 (0.03) & 15.54 (0.03) & 14.86 (0.03) & 14.27 (0.04) & 14.29 (0.05) \\
 8 & 19.61 (0.05) & 19.15 (0.06) & 18.89 (0.04) & 18.51 (0.04) & \nodata & \nodata & \nodata		   \\
 9 & 20.63 (0.04) & 19.19 (0.04) & 18.35 (0.04) & 17.18 (0.04) & 15.94 (0.06) & 15.32 (0.07) & 15.33 (0.13) \\
10 & 19.97 (0.06) & 18.68 (0.03) & 17.87 (0.03) & 17.05 (0.02) & 15.96 (0.08) & 15.40 (0.07) & 15.29 (0.13) \\
12 & 17.08 (0.03) & 16.46 (0.02) & 16.12 (0.04) & 15.73 (0.02) & 15.26 (0.04) & 14.84 (0.05) & 14.76 (0.09) \\
12 & 20.12 (0.04) & 19.18 (0.05) & 18.62 (0.05) & 18.12 (0.07) &  \nodata & \nodata & \nodata		   \\
13 & 20.97 (0.03) & 19.58 (0.01) & 18.76 (0.02) & 17.79 (0.04) & 16.65 (0.11) & 16.20 (0.15) & 15.38 (0.15)  \\
14 & 18.99 (0.04) & 18.47 (0.04) & 18.11 (0.04) & 17.80 (0.03) &  \nodata & \nodata & \nodata		   \\
15 & 15.99 (0.04) & 14.98 (0.03) & 14.38 (0.06) & 13.83 (0.05) & 12.93 (0.03) & 12.41 (0.02) & 12.33 (0.03) \\
\enddata
\tablenotetext{a}{Quoted uncertainties are $1\sigma$.}
\end{deluxetable}

\clearpage
\startlongtable
\begin{deluxetable}{@{}lccccccccl@{}}
\tablenum{3}
\tablewidth{0pt}
\tablecolumns{9}
\tablecaption{Optical Johnson-Cousins Photometry of SN~2010bt (\textsc{vegamag})\tablenotemark{a}
\label{table_JCph}}
\tablehead{
\colhead{Date} & \colhead{MJD} & \colhead{Phase\tablenotemark{b}} & \colhead{$U$} & \colhead{$B$} & \colhead{$V$} & \colhead{$R$}  & \colhead{$I$} & \colhead{Instrument key} \\ 
\colhead{} & \colhead{} & \colhead{(days)} & \colhead{(mag)} & \colhead{(mag)} & \colhead{(mag)} & \colhead{(mag)} & \colhead{(mag)} & \colhead{}}
\startdata
20010813 & 52134.0 & -3169.1 & \nodata &  \nodata &  \nodata &  $>  18.0 $&  \nodata & JAG  \\ 
20100417 & 55303.1 &     0.0 & \nodata &  \nodata &  \nodata &  16.00 (0.41) &  \nodata & BO  \\ 
20100418 & 55304.1 &     1.0 & \nodata &  \nodata &  \nodata &  16.08 (0.18) &  \nodata & BO  \\ 
20100418 & 55304.4 &     1.3 & \nodata &  17.00 (0.08) &  16.33 (0.08) &  16.05 (0.10) &  15.79 (0.12) & EFOSC2  \\ 
20100420 & 55306.4 &     3.3 & \nodata &  17.11 (0.03) &  16.54 (0.03) &  16.26 (0.04) &  15.88 (0.04) & ANDICAM  \\ 
20100420 & 55306.9 &     3.8 & 16.88 (0.14) &  17.33 (0.15) &  16.58 (0.17) &  \nodata &  \nodata & {\it Swift}  \\ 
20100422 & 55308.1 &     5.0 & 17.09 (0.17) &  17.35 (0.16) &  16.42 (0.15) &  \nodata &  \nodata & {\it Swift}  \\ 
20100423 & 55309.4 &     6.2 & \nodata &  17.42 (0.08) &  16.81 (0.07) &  16.50 (0.05) &  16.17 (0.04) & ANDICAM  \\ 
20100424 & 55310.2 &     7.0 & 17.26 (0.19) &  17.58 (0.18) &  17.09 (0.24) &  \nodata &  \nodata & {\it Swift}  \\ 
20100426 & 55312.8 &     9.7 & 17.71 (0.26) &  17.75 (0.22) &  \nodata &  \nodata &  \nodata & {\it Swift} \\ 
20100427 & 55313.3 &    10.2 & \nodata &  17.89 (0.18) &  17.16 (0.16) &  16.73 (0.09) &  16.39 (0.14) & ANDICAM  \\ 
20100428 & 55314.7 &    11.6 & 17.71 (0.26) &  17.77 (0.20) &  17.25 (0.26) &  \nodata &  \nodata & {\it Swift}  \\ 
20100430 & 55316.0 &    12.9 & $>  18.3 $&  \nodata &  \nodata &  \nodata &  \nodata & {\it Swift}  \\ 
20100430 & 55316.0 &    12.9 & \nodata &  18.23 (0.28) &  17.46 (0.30) &  \nodata &  \nodata & {\it Swift}  \\ 
20100430 & 55316.4 &    13.3 & \nodata &  18.11 (0.15) &  17.36 (0.13) &  16.89 (0.10) &  16.66 (0.09) & ANDICAM  \\ 
20100501 & 55317.9 &    14.7 & \nodata &  \nodata &  $>  17.8 $&  \nodata &  \nodata & {\it Swift}  \\ 
20100502 & 55318.9 &    15.7 & 18.15 (0.35) &  18.16 (0.27) &  \nodata &  \nodata &  \nodata & {\it Swift}  \\ 
20100503 & 55319.4 &    16.3 & \nodata &  18.24 (0.14) &  17.52 (0.12) &  16.95 (0.09) &  16.70 (0.10) & ANDICAM  \\ 
20100504 & 55320.9 &    17.8 & 18.27 (0.39) &  18.46 (0.34) &  17.69 (0.38) &  \nodata &  \nodata & {\it Swift}  \\ 
20100506 & 55322.1 &    19.0 & \nodata &  \nodata &  \nodata &  17.18 (0.14) &  \nodata & BO  \\ 
20100506 & 55322.2 &    19.1 & 18.10 (0.35) &  18.27 (0.30) &  \nodata &  \nodata &  \nodata & {\it Swift} \\ 
20100506 & 55322.2 &    19.1 & \nodata &  \nodata &  $>  17.7 $&  \nodata &  \nodata & {\it Swift}  \\ 
20100507 & 55323.4 &    20.2 & \nodata &  18.56 (0.06) &  17.79 (0.05) &  17.30 (0.04) &  16.96 (0.06) & ANDICAM  \\ 
20100508 & 55324.5 &    21.4 & $>  18.3 $&  $>  18.6 $&  \nodata &  \nodata &  \nodata & {\it Swift} \\ 
20100510 & 55326.3 &    23.2 & \nodata &  18.98 (0.13) &  17.98 (0.11) &  17.52 (0.06) &  17.22 (0.06) & ANDICAM  \\ 
20100511 & 55327.1 &    24.0 & \nodata &  \nodata &  \nodata &  $>  16.7 $&  \nodata & BO  \\ 
20100518 & 55334.4 &    31.3 & \nodata &  19.58 (0.11) &  18.59 (0.09) &  17.98 (0.06) &  17.69 (0.07) & ANDICAM  \\ 
20100519 & 55335.4 &    32.3 & \nodata &  19.60 (0.08) &  18.62 (0.07) &  18.01 (0.06) &  17.68 (0.06) & ANDICAM  \\ 
20100523 & 55339.3 &    36.2 & \nodata &  19.84 (0.14) &  18.84 (0.12) &  18.19 (0.08) &  17.91 (0.12) & ANDICAM  \\ 
20100524 & 55340.1 &    37.0 & \nodata &  \nodata &  \nodata &  $>  16.5 $&  \nodata & BO  \\ 
20100524 & 55340.3 &    37.2 & \nodata &  19.89 (0.14) &  18.74 (0.12) &  18.23 (0.11) &  17.95 (0.12) & ANDICAM  \\ 
20100526 & 55342.3 &    39.2 & \nodata &  20.10 (0.12) &  19.04 (0.11) &  18.34 (0.07) &  18.03 (0.08) & ANDICAM  \\ 
20100531 & 55347.4 &    44.2 & \nodata &  20.47 (0.13) &  19.22 (0.12) &  18.48 (0.07) &  18.16 (0.09) & ANDICAM  \\ 
20100604 & 55351.3 &    48.2 & \nodata &  20.79 (0.13) &  19.48 (0.11) &  18.63 (0.10) &  18.24 (0.07) & ANDICAM  \\ 
20100613 & 55360.1 &    57.0 & \nodata &  \nodata &  \nodata &  $>  16.6 $&  \nodata & BO  \\ 
20100915 & 55454.2 &   151.1 & \nodata &  $>  19.8 $&  $>  19.8 $&  $>  18.9 $&  $>  19.7 $& EFOSC2  \\ 
20101004 & 55473.1 &   170.0 & \nodata &  $>  20.4 $&  $>  20.0 $&  $>  19.3 $&  $>  18.7 $& EFOSC2  \\ 
20101009 & 55478.4 &   175.3 & \nodata &  \nodata &  22.50 (0.05) &  \nodata &  \nodata & HST\_ACS/WFC \\ 
20101028 & 55497.2 &   194.1 & \nodata &  $>  21.3 $&  $>  21.0 $&  $>  20.2 $&  $>  19.7 $& EFOSC2  \\ 
20150917 & 57283.0 &   1979.9 & \nodata &  $>  18.6 $ & \nodata & $>  18.5 $ & \nodata & LDSS-3 \\
20151014 & 57309.0 &   2005.9 & \nodata &  \nodata &  $> 24.4$ &  \nodata &  \nodata & HST\_WFC3\\
\enddata
\tablenotetext{a}{Quoted uncertainties are $1\sigma$.}
\tablenotetext{b}{Phases are relative to the discovery, MJD = 55303.1.}
\end{deluxetable}

\begin{deluxetable}{@{}ccccccc@{}}
\tablenum{4}
\tablewidth{0pt}
\tablecolumns{7}
\tablecaption{Near-Infrared Photometry of SN~2010bt (\textsc{vegamag})\tablenotemark{a}
\label{table_NIRph}}
\tablehead{
\colhead{Date} & \colhead{MJD} & \colhead{Phase\tablenotemark{b}} & \colhead{$J$} & \colhead{$H$} & \colhead{$K$} & \colhead{Instrument key} \\ 
\colhead{} & \colhead{} & \colhead{(days)} & \colhead{(mag)} & \colhead{(mag)} & \colhead{(mag)} & \colhead{}}
\startdata
20100525 & 55341.0 &    37.9 & \nodata &  \nodata &  16.69 (0.23) & HAWK-I  \\ 
20100606 & 55353.0 &    49.9 & \nodata &  \nodata &  17.51 (0.31) & HAWK-I  \\ 
20100707 & 55384.0 &    80.9 & \nodata &  \nodata &  18.49 (0.22) & HAWK-I  \\ 
20100917 & 55456.1 &   153.0 & $>  18.7 $&  $>  17.4 $&  $>  18.9 $& SOFI  \\ 
20101029 & 55498.1 &   195.0 & $> 17.2$ & $> 15.4$ & \nodata & SOFI\\ 
\enddata
\tablenotetext{a}{Quoted uncertainties are $1\sigma$.}
\tablenotetext{b}{Phases are relative to the discovery, MJD = 55303.1.}
\end{deluxetable}

\begin{deluxetable}{@{}cccccc@{}}
\tablenum{5}
\tablewidth{0pt}
\tablecolumns{7}
\tablecaption{{\it Swift\/} UV Photometry of SN~2010bt (\textsc{vegamag})\tablenotemark{a}
\label{table_UVph}}
\tablehead{
\colhead{Date} & \colhead{MJD} & \colhead{Phase\tablenotemark{b}} & \colhead{UVW1} & \colhead{UVM2} & \colhead{UVW2} \\ 
\colhead{} & \colhead{} & \colhead{(days)} & \colhead{(mag)} & \colhead{(mag)} & \colhead{(mag)}}
\startdata
20100420 & 55306.9 &     3.8 & $>  18.5 $&  $>  18.8 $&  $>  18.9 $ \\ 
20100422 & 55308.1 &     5.0 & 18.20 (0.34) &  \nodata &  \nodata   \\ 
20100422 & 55308.2 &     5.0 & \nodata &  $>  18.8 $&  $>  18.9 $ \\ 
20100424 & 55310.2 &     7.0 & 18.48 (0.42) &  \nodata &  \nodata   \\ 
20100424 & 55310.2 &     7.0 & \nodata &  $>  18.7 $&  $>  18.9 $ \\ 
20100426 & 55312.8 &     9.7 & $>  18.5 $&  \nodata &  $>  18.4 $ \\ 
20100428 & 55314.7 &    11.6 & $>  18.5 $&  $>  18.3 $&  $>  18.9 $ \\ 
20100430 & 55316.0 &    12.9 & $>  18.6 $&  $>  18.9 $&  $>  19.0 $ \\ 
20100502 & 55318.9 &    15.7 & $>  18.6 $&  $>  18.9 $&  $>  19.0 $ \\ 
20100504 & 55320.9 &    17.8 & $>  18.5 $&  $>  18.8 $&  $>  18.9 $ \\ 
20100506 & 55322.2 &    19.1 & $>  18.5 $&  $>  18.9 $&  $>  18.9 $ \\ 
20100508 & 55324.5 &    21.4 & $>  18.5 $&  $>  18.9 $&  $>  18.9 $ \\ 
\enddata
\tablenotetext{a}{Quoted uncertainties are $1\sigma$.}
\tablenotetext{b}{Phases are relative to the discovery, MJD = 55303.1.}
\end{deluxetable}

\begin{deluxetable}{@{}llccccc@{}}
\tablenum{6}
\tablewidth{0pt}
\tablecolumns{7}
\tablecaption{Properties of the Comparison Supernovae
\label{tabla_SNe}}
\tablehead{
\colhead{SN} & \colhead{Host galaxy} & \colhead{Redshift} & \colhead{Distance$^{\dagger}$} & \colhead{$E(B-V)_{\rm tot}$}  & \colhead{$V_{\rm max}$ Date} & \colhead{Source} \\
\colhead{} & \colhead{} & \colhead{} & \colhead{(Mpc)} & \colhead{(mag)} & \colhead{(MJD)} & \colhead{} }
\startdata
1996al & NGC~7689 & 0.007 &22.9 & 0.110 & 50265.0 & a \\
1998S & NGC~3877 & 0.002 &15.7 & 0.219 & 50890.5 & b \\
2009ip & NGC~7259 & 0.006 & 25.0 & 0.019 & 56203.5 &  c \\
2015bh & NGC~2770 & 0.007 & 29.3 & 0.208 & 57167.0 & d \\
2010bt & NGC~7130 & 0.016 & 65.4 (4.6) & 0.40 (0.14) & 55293.1$^{\ddagger}$ & This work \\
\enddata
$^{\dagger}${Distances have been scaled to H$_0 = 73$ \kms\,Mpc$^{-1}$.}\\
\noindent $^\ddagger$We have assumed $V$ maximum date = discovery date (MJD 55303.1) $- 10$ d.
\noindent Sources: a, \citet{benetti16}; b, \citet{fassia00}, \citet{leonard00}, \citet{pozzo04}, NED; c, \citet{pastorello13}, \citet{fraser13}, \citet{mauerhan13a}, \citet{margutti14}; d, \citet{eliasrosa16}.
\end{deluxetable}

%
\subsection{Spectroscopy}\label{SNspec}

Four low-resolution optical spectra of SN~2010bt were obtained on 2010 April 18 (classification spectrum), September 15 and 16, and October 28 with the 3.58~m NTT+EFOSC2 at ESO of La Silla (Chile). Only the first spectrum was taken with the slit along the parallactic angle to avoid differential flux losses (\citealt{filippenko82}). For the other spectra it was necessary to point to a nearby bright star, and then rotate the slit to position the SN~2010bt site inside the aperture. Basic information on our spectra is reported in Table \ref{table_spectra}.

All spectra were reduced following standard procedures with {\sc iraf} routines. The two-dimensional (2D) spectroscopic frames were debiased and flat-fielded, before the optimized extraction \citep{horne86} of the 1D spectra. Wavelength calibration was accomplished with the help of arc-lamp exposures obtained the same night. Small adjustments estimated from night-sky lines in the object frames were applied. The spectra were flux calibrated using the well-exposed continua of spectrophotometric standard stars \citep{oke90,hamuy92,hamuy94}. An atmospheric extinction correction was applied using tabulated extinction coefficients for the ESO-La Silla Observatory. The strongest telluric absorption bands were removed using the standard-star spectra. Finally, the flux of each spectrum was cross-checked against the photometry. 

\begin{figure}
\figurenum{6}
\plotone{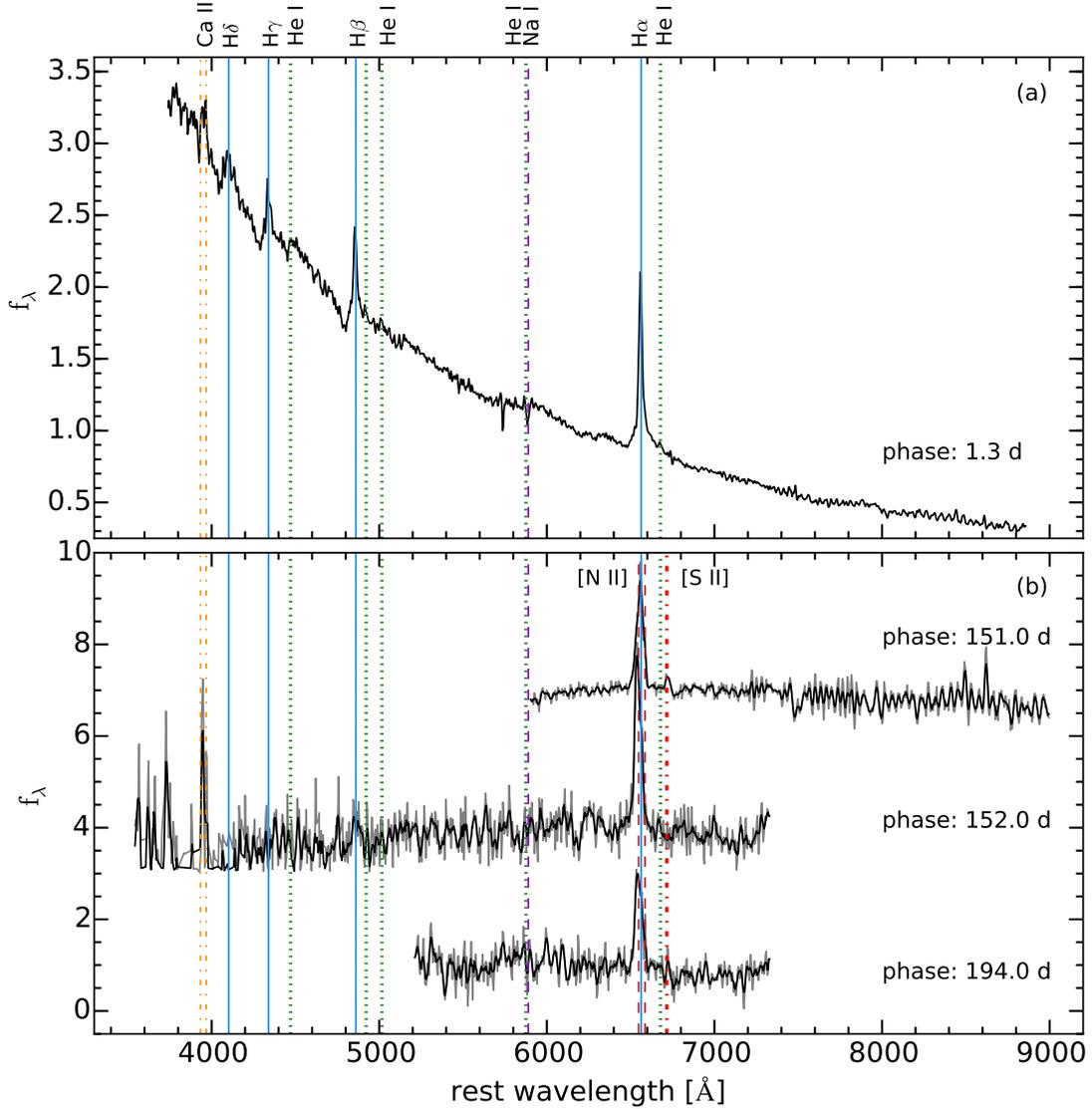}
\epsscale{0.8}
\caption{(a) Early-time and (b) late-time optical spectra of SN~2010bt. The late spectra are shown in grey, with a boxcar-smoothed (using a 5-pixel window) version of the spectra overplotted in black. The locations of the most prominent spectral features are indicated by vertical lines.\label{fig_spectrum}}
\end{figure}

\begin{figure}
\figurenum{7}
\plotone{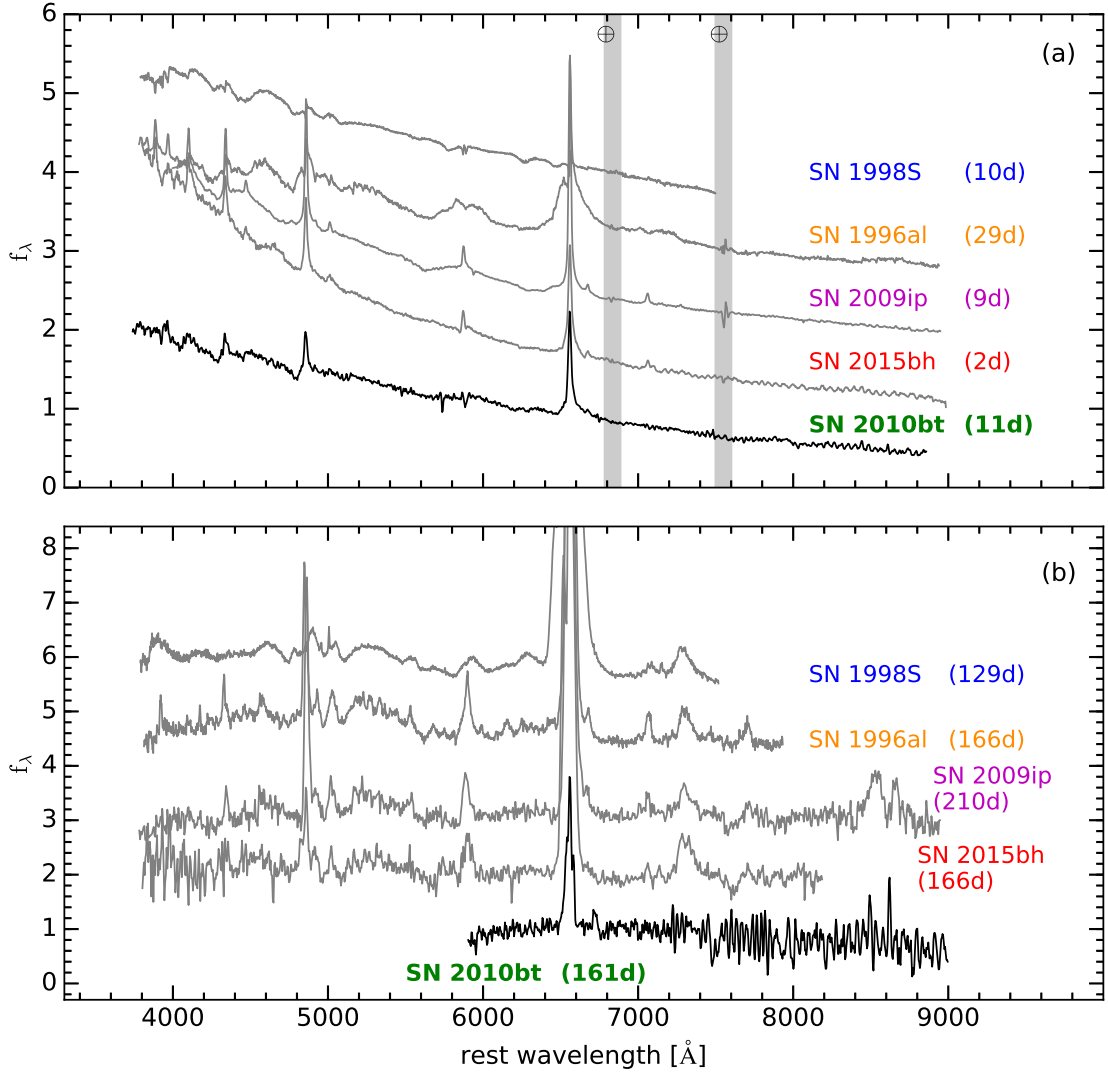}
\epsscale{0.8}
\caption{Comparison of SN~2010bt (a) early-time and (b) late-time optical spectra, along with those of the interacting SNe~1996al, 1998S, 2009ip, and 2015bh at similar epochs. All spectra have been corrected for their host-galaxy recession velocities and for extinction (values adopted from the literature; see also Table \ref{tabla_SNe}). Ages are relative to $V$ maximum light.  \label{fig_spectrumcomp}}
\end{figure}

\begin{figure}
\figurenum{8}
\plotone{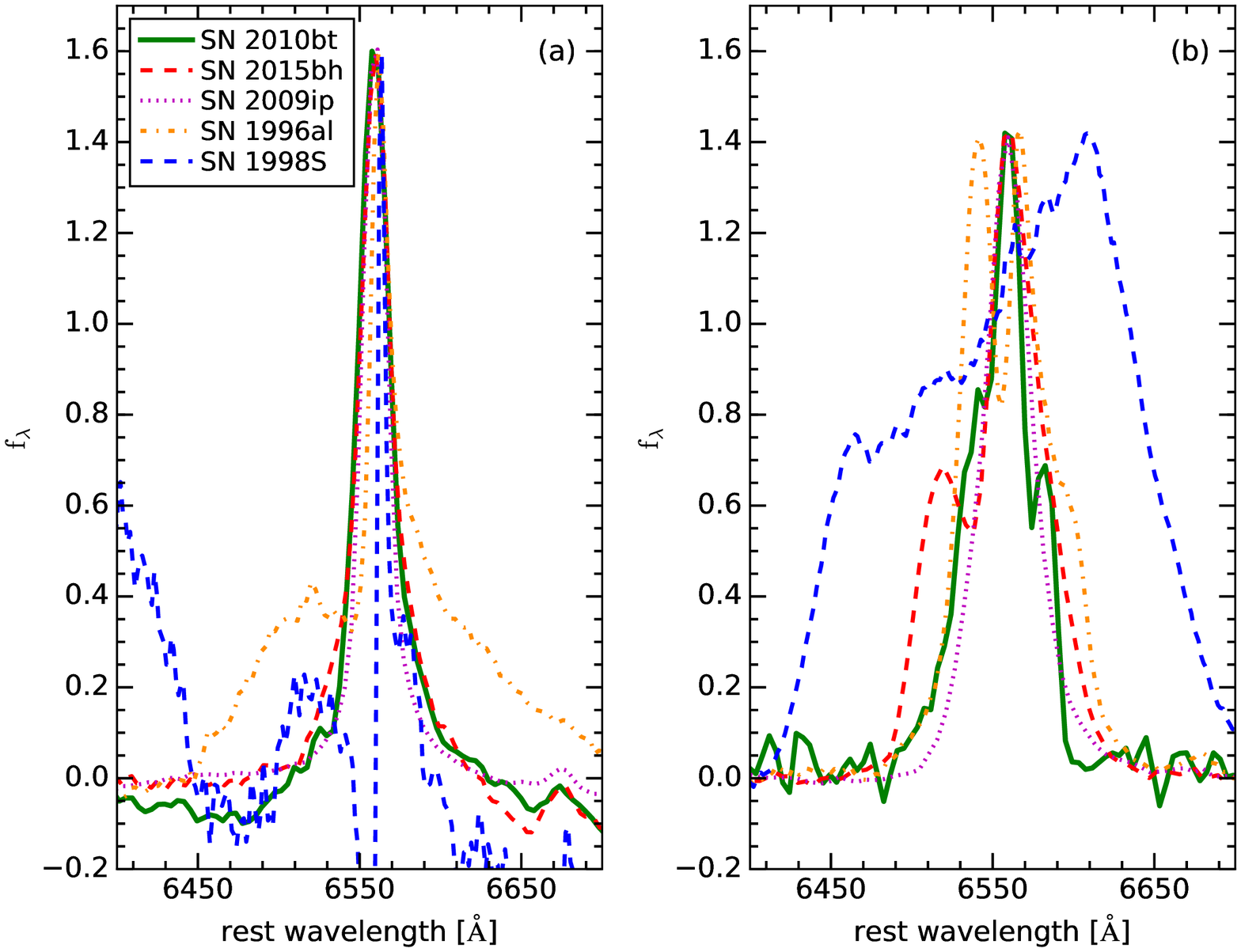}
\epsscale{0.8}
\caption{Close-up view of the H${\alpha}$ profiles of the spectra in Figure \ref{fig_spectrumcomp}. Panel (a) shows the early-time optical spectra, and panel (b) those taken at late times. The profiles are normalized to the same peak flux.\label{fig_spectrumcomp_Halphazoom}}
\end{figure}

Figure \ref{fig_spectrum} shows the sequence of optical spectra of SN~2010bt\footnote{Our spectra are available on WiseREP \citep{yaron12}.}, and in Figures \ref{fig_spectrumcomp} and \ref{fig_spectrumcomp_Halphazoom} we compare some of these spectra with those of SNe~IIn~1996al \citep{benetti16}, 1998S \citep{leonard00}, and the transients SN~2009ip \citep{pastorello13,fraser13} and SN~2015bh \citep{eliasrosa16} at similar epochs. All of the spectra have been corrected for extinction and deredshifted using values from the literature (see also Table \ref{tabla_SNe}).

The early-time spectrum (Figure \ref{fig_spectrum}, top panel) exhibits a blue continuum, with relatively weak spectral features, except for the strong Balmer emission lines with complex, yet relatively narrow profiles. SN~2010bt does not show signs of the blue pseudocontinuum that characterizes some of the most energetic SNe~IIn such as SN~1988Z \citep{turatto93,kiewe12}. The pseudocontinuum is a ``blue excess" arising from the overlap of emission lines when the expanding ejecta of the SN interact with the surrounding circumstellar material. The lack of a pseudocontinuum indicates that for SN~2010bt, the continuum is thermal with a blackbody-like behavior.

Balmer emission lines show P-Cygni absorption components with expansion velocities of 4000--3500 \kms, estimated from their absorption minima. A blend of He\,{\sc i} $\lambda$5876 and Na\,{\sc i}~D is also visible in the spectrum, with the broad emission centered at $\sim 5950$~\AA\ and with a very broad FWHM $\approx 290$~\AA. The photospheric temperature at this epoch, estimated by fitting the SED of SN~2010bt with a blackbody function, is around 12,900~K (see also Table \ref{table_specpar}). Given this temperature and the pseudobolometric luminosity of SN~2010bt at that epoch, a radius of the photosphere of $\sim 8 \times 10^{14}$~cm is inferred. 

Consistent with the photometry, the H$\alpha$ profile of SN~2010bt resembles those of the interacting transients SNe~2009ip and 2015bh, and shows differences with SNe~1996al and 1998S (see the close-up view of the H$\alpha$ profiles in Figures \ref{fig_spectrumcomp} and \ref{fig_spectrumcomp_Halphazoom}). The blackbody temperature shaping the continuum, though, is similar to that of SN~1998S, as are the strengths of the He\,{\sc i} lines.

The extractions of the late-time spectra ($> 150$ d) were arduous, given their low signal-to-noise ratio (S/N). Note that these spectra are contemporaneous with the upper photometric limits measured for SN~2010bt. The only distinct features are those of H$\beta$, H$\alpha$, together with typical residual emission lines, such as [N\,{\sc ii}] $\lambda\lambda$6548, 6584 and [S\,{\sc ii}] $\lambda\lambda$6717, 6731, from neighboring H\,{\sc ii} regions. 
Despite the low S/N of the late-time SN~2010bt spectra, we notice some differences comparing its H$\alpha$ profile to that of the other SNe. None of the transients share the same profile shape, with that of SN~2010bt being double peaked, yet narrower than that of the other objects (see the H$\alpha$ profiles magnified in Figure \ref{fig_spectrumcomp_Halphazoom}). 

\begin{deluxetable}{@{}ccrcccc@{}}
\tablenum{7}
\tablewidth{0pt}
\tablecolumns{7}
\tablecaption{Log of Spectroscopic Observations of SN~2010bt \label{table_spectra}}
\tablehead{
\colhead{UT Date} & \colhead{MJD} & \colhead{Phase\tablenotemark{a}} & \colhead{Instrument} & \colhead{Grism/grating + slit} & \colhead{Spectral range} & \colhead{Resolution} \\ 
\colhead{} & \colhead{} & \colhead{(days)} & \colhead{key} & \colhead{} & \colhead{(\AA)} & \colhead{(\AA)}}
\startdata
20100418 & 55304.4  & 1.3 & EFOSC2 & gm11+gm16+1.5$''$ & 3400--10,300 & 14 \\  
20100915 & 55454.1  & 151.0 & EFOSC2 & gm11+gm16+1.0$''$ & 3400--10,300 & 14 \\  
20100916 & 55455.1  & 152.0 & EFOSC2 & gm11+1.0$''$ & 3400--7500 & 14 \\  
20101028 & 55497.1  & 194.0 & EFOSC2 & gm11+1.0$''$ & 3400--7500 & 14 \\  
\enddata
\tablenotetext{a}{Phases are relative to the date of discovery, MJD = 55303.1.}
\end{deluxetable}



%
\subsubsection{Hydrogen Feature Decomposition}\label{SNspecdecomp}

The H${\alpha}$ line profile of SN~2010bt appears to consist of multiple components with evident changes in morphology between early and late phases. We have decomposed the line into a Lorentzian profile for the narrow component and two Gaussian functions for the emission and absorption components of the broader P-Cygni profile at the early epochs, and into two Lorentzian profiles for the oldest phases. To do this, we have used a Python script for least-squares minimization. 

\begin{figure}
\figurenum{9}
\plotone{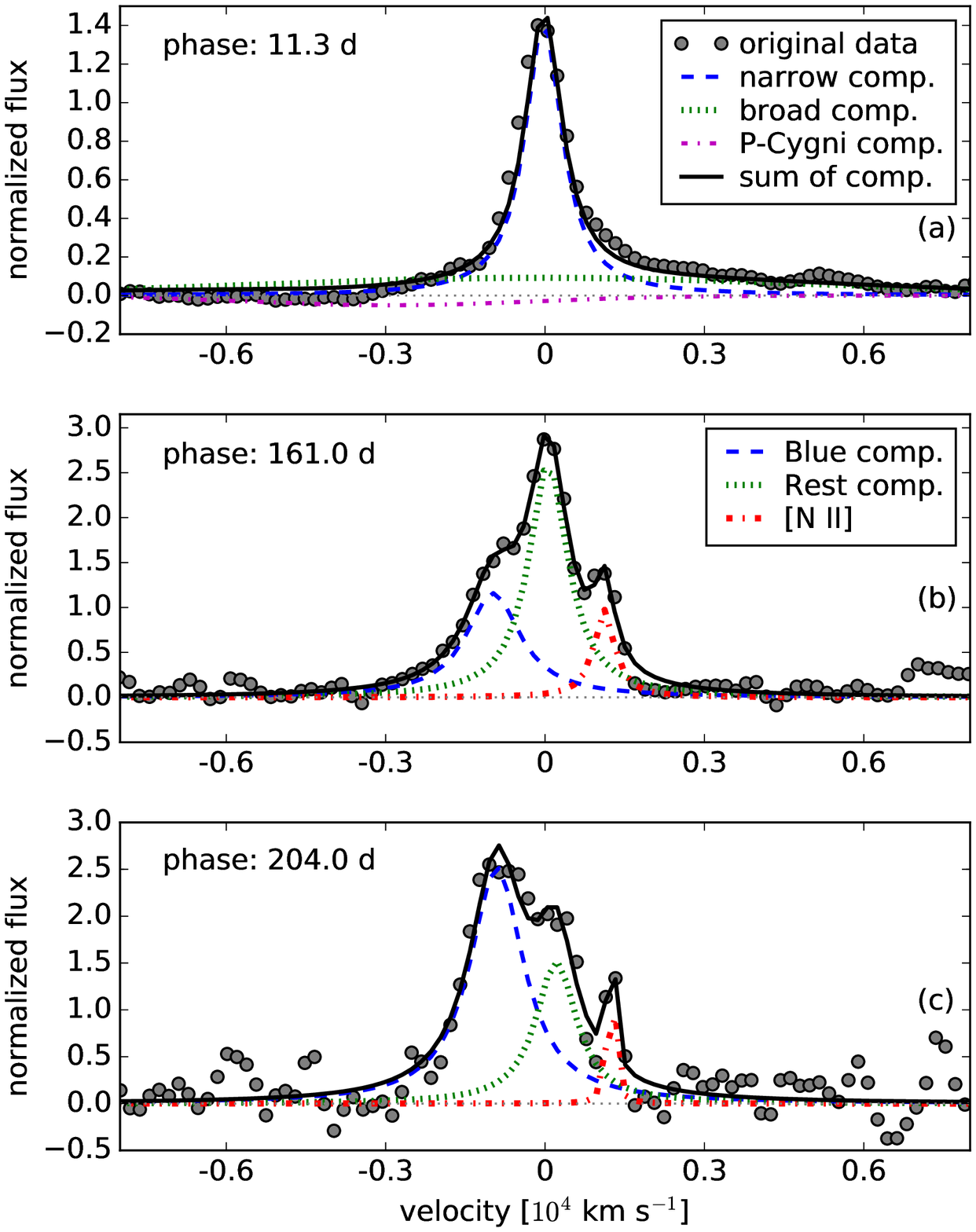}
\epsscale{0.5}
\caption{Decomposition of the H${\alpha}$ emission line of SN~2010bt at phases 11.3, 161.0, and 204.0 d from the assumed $V$-maximum date. 
\label{fig_specdecomp}}
\end{figure}

Figure \ref{fig_specdecomp} presents the H${\alpha}$ emission-line decomposition for three of our four epochs. The procedure independently fits the parameters, while the uncertainties were estimated using a bootstrap resampling technique, varying randomly the flux of each pixel according to a normal distribution having variance equal to the noise of the continuum. The velocities of the different gas components are listed in Table \ref{table_specpar}. 

The relatively narrow H$\alpha$ emission at the first epoch (phase $\sim 11.3$ d) is resolved with a FWHM of $\sim 750$ \kms\ (after correction for instrumental resolution), while the broader component has FWHM $\approx 12,800$ \kms. In addition, the blue side of the H${\alpha}$ profile is ``absorbed" by a P-Cygni component with minimum at $\sim 3000$ \kms. The broader component is redshifted, probably caused by electron scattering as the H${\alpha}$ line photons diffuse through dense CSM ahead of the shock. The observed luminosity of H${\alpha}$ was estimated from the integrated flux of the entire line to be $5 \times 10^{40}$ erg s$^{-1}$. 

\begin{figure}
\figurenum{10}
\epsscale{0.6}
\plotone{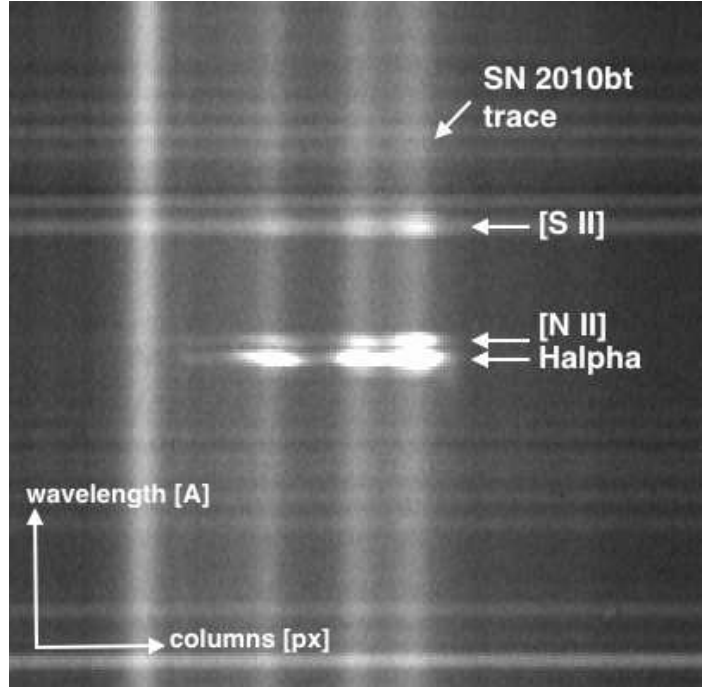}
\caption{The H${\alpha}$ emission line in the 2D spectrum of SN~2010bt taken with NTT+EFOSC2 on 2010 September 15 (phase 161.0~d from the assumed $V$-maximum date). 
\label{fig_spec2D}}
\end{figure}

At later phases ($>150$ d), the H${\alpha}$ profile is well reproduced with two Lorentzian components called ``Blue'' and ``Rest,'' given that they are centered at wavelengths of $\sim 6542$ \AA\ and $\sim 6565$ \AA, respectively. Clearly visible is a broad feature of H${\alpha}$ in the 2D images of the late-time SN~2010bt spectra (see Figure \ref{fig_spec2D}). It is also evident that these spectra are likely contaminated by neighboring H\,{\sc ii} regions. In fact, a residual emission line centered at $\sim 6588$ \AA, most likely corresponding to [N\,{\sc ii}] $\lambda6584$, is present. Consequently, we have considered this line in the decomposition of H${\alpha}$ by adding a third Lorentzian component. The FWHMs of the Blue and Rest components at all late epochs are relatively constant, at $\sim 1500$ \kms\ and $\sim 1000$ \kms, respectively. The total luminosity of the H$\alpha$ line at these phases decreases to $3 \times 10^{39}$ erg s$^{-1}$. 

\begin{deluxetable}{cccccccccc}
\tablenum{8}
\tablewidth{0pt}
\tablecolumns{9}
\setlength{\tabcolsep}{0.02in}
\tablecaption{Main Parameters Inferred from Spectra of SN~2010bt \label{table_specpar}}
\tablehead{
\colhead{UT Date} & \colhead{MJD} & \colhead{Phase\tablenotemark{a}} & \colhead{Temp.\tablenotemark{b}} & 
\colhead{Radius\tablenotemark{c}} & \colhead{FWHM$_\mathrm{H\alpha,nar}$\tablenotemark{d}} & \colhead{FWHM$_\mathrm{H\alpha,br}$} & \colhead{$v_{\rm P-Cyg}$} & \colhead{$L({\rm H}\alpha)$}  \\
\colhead{} & \colhead{} & \colhead{(days)} & \colhead{(K)} & \colhead{($10^{14}$ cm)} & \colhead{(\kms)} &
\colhead{(\kms)} & \colhead{(\kms)} & \colhead{($10^{39}$ erg s$^{-1}$)} }
\startdata
20100418 & 55304.4  & 11.3 & 12900 & 8 (2) & 750 (100) & 12,800 (1600) & 3000 (800) & 50 (7) \\
\hline
 & &  & & & FWHM$_\mathrm{Blue}$ & FWHM$_\mathrm{Rest}$ & & \\
 & &  & & & (\kms) & (\kms) &  & \\
\hline
20100915 & 55454.1  & 161.0 & \nodata & \nodata & 1500 (400) & 1050 (300) & \nodata & 3 (1) \\
20100916 & 55455.1  & 162.0 & \nodata & \nodata & 1550 (400) & 750 (500)  & \nodata & 2 (1) \\
20101028 & 55497.1  & 204.0 & \nodata & \nodata & 1450 (400) & 1100 (350) & \nodata & 1 (0.4) \\
\enddata
\tablenotetext{a}{Phases are relative to the assumed $V$ maximum date = discovery date (MJD 55303.1) $- 10$ d.}
\tablenotetext{b}{We consider a conservative uncertainty in the temperature of about $\pm 500$~K.}
\tablenotetext{c}{We have propagated the uncertainties from the Stefan-Boltzmann equation.}
\tablenotetext{d}{The velocities are computed from the decomposition of the H${\alpha}$ profile.}
\end{deluxetable}

%
\subsection{Explosion Date of SN~2010bt}\label{SNexpl}

\citet{monard10} reported that no sources were visible at the SN position in images taken on 2009 Dec. 22.80 (limit $> 17.8$ mag, or $M_R \gtrsim -17.3$ mag, for this particular case). Nothing was also visible in a HAWK-I image taken on 2009 July 26.60 (limit $K > 19.0$ mag; in this case $M_K \gtrsim -15.2$ mag; \citealt{miluzio13}). Therefore, the SN explosion occurred less than 115 d before discovery. 

A more accurate date of the explosion is difficult to estimate in the case of SN~2010bt. The main complications are the peculiar behaviors of both its photometric and spectroscopic evolution. 

The light curves (see Section~\ref{SNph}) show that the SN was discovered after maximum light. Assuming that the $V$-band light curve of SN~2010bt has a similar decay slope as that of SN~2009ip (see Section \ref{SNph} and Figure \ref{fig_abs}), the former would have been discovered no more than $10$ days after maximum light. This estimate is also confirmed by the behavior of the color curves. Considering that  SNe~IIn generally have a rise time of $> 5$ d \citep{ofek14}, SN~2010bt would have exploded {\it at least 15 d} before the discovery. On the other hand, and according to the template-fitting code {\sc GELATO}, acceptable matches with SNe~IIn at phases between {\it 10 and 50 d} after explosion are found for the early-time SN~2010bt spectrum (although the best fit is at 10 d).

Thus, in view of all these uncertainties, we are able to determine only that the explosion occurred {\it 15--50 d before discovery}.
%
\section{Identification of the Progenitor Candidate}\label{identification}

In order to search for a possible SN progenitor, we isolated archival\footnote{\url{http://archive.stsci.edu/hst/}} {\sl HST\/} images of NGC~7130 taken with the Wide-Field and Planetary Camera 2 (WFPC2) in F606W ($\sim V$; 500~s exposure) on 1994 August 23 by program GO-5479 (PI M.~Malkan), and with the High Resolution Channel (HRC) of the Advanced Camera for Surveys (ACS) in F330W ($\sim u$; 1200~s exposure) on 2003 May 08 by program GO-9379 (PI H.~Schmitt). We worked with drizzled images downloaded from the Hubble Legacy Archive\footnote{\url{http://hla.stsci.edu/hlaview.html}}. These images have been resampled to a uniform grid to correct for geometric distortions. 

We performed relative astrometry by geometrically transforming the pre-explosion images to match those taken after the explosion. We first used the ground-based, post-explosion CTIO/SMARTS image taken on 2010 April 23 (with seeing $1.5''$) to approximately locate the position of the SN in the WFPC2 images. Then, we confirmed the identification of this candidate through high-resolution {\sl HST\/} ACS images using the Wide Field Channel (WFC, $\sim 0.05$\arcpx; a pair of observations of 40~s and 80~s exposures), obtained through the F606W filter on 2010 October 09, as part of our Target-of-Opportunity program GO-11575 (PI S.~Van Dyk). The individual exposures of this trigger were drizzled to produce a final mosaic. However, as we showed in Section \ref{SNph}, the SN was already quite faint at the time of these observations ($F606W (\sim V) = 22.5$ mag; see Table~\ref{table_JCph}), leaving some ambiguity in its identification. Therefore, we used the {\it HST} post-explosion image to identify the progenitor in the pre-SN WFPC2 and ACS/HRC images, as follows.

(i) Initially, we obtained a precise SN position, at $\alpha = 21^{\rm h} 48^{\rm m} 20{\fs}28$, $\delta = -34\degr 57\arcmin 17{\fdg}7$ (J2000.0), with root-mean-squared (RMS) uncertainties of $\sim 0{\farcs}1$, based on 9 point-like sources as fiducials in the $V$-band NTT+EFOSC2 image taken on 2010 April 18. We measured the centroids of the fiducial sources with the {\sc iraf} tasks {\sc imexamine}. We adopted the 2MASS Point Source Catalog as the astrometric grid and used the {\sc iraf} task {\sc ccmap} to obtain the solution. At that point, we were able to confirm the SN position in the {\it HST} post-explosion mosaic, with RMS uncertainty $< 0.06''$, and located the faint SN.

(ii) Then, we achieved high-precision relative astrometry between the pre-explosion WFPC2 F606W (we worked only with the drizzled PC image, since the SN site is located on this chip, with $0{\farcs}045$ pixel$^{-1}$) and ACS/HRC F330W images (with $\sim 0{\farcs}025$ pixel$^{-1}$), and the post-explosion ACS/WFC image, by geometrically transforming the former pair to match the latter. We used 17 point-like sources in common between the three sets of images and the {\sc iraf} task {\sc geomap} for the transformation.

\begin{figure}
\figurenum{11}
\epsscale{1.2}
\plotone{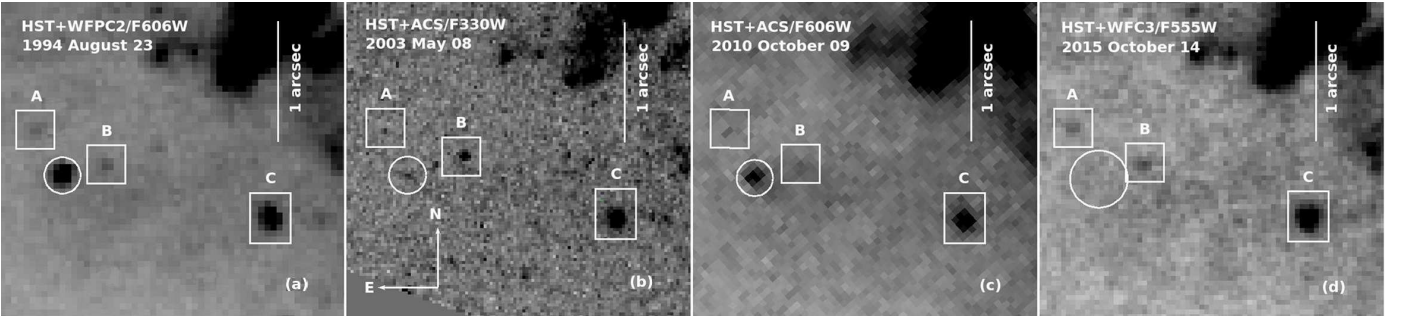}
\caption{Subsections of the NGC~7130 pre-explosion (a) {\sl HST\/} WFPC2 image in F606W  and (b) {\sl HST\/} ACS/HRC image in F330W, along with (c) the post-explosion {\sl HST\/} ACS/WFC image of the SN site in F606W and (d) the late-time {\sl HST\/} WFC3/UVIS image in F555W. The positions of the SN candidate progenitor and SN are indicated by {\it circles}, each with a radius of 3 pixels (between $0{\farcs}08$ and $0{\farcs}15$), except in panel (d), for which the radius is 6 pixels ($\sim 0{\farcs}23$). The positions of three neighboring sources of SN~2010bt (``A,'' ``B,'' and ``C'') are also indicated. \label{fig_progenitor}}
\end{figure}

The positions (and their uncertainties) for the SN and the progenitor candidate are obtained by averaging the measurements from two centroiding methods, the task {\sc daofind\/} within {\sc iraf}/{\sc daophot} and {\sc imexamine} within {\sc iraf}. As a result, in the pre-SN F606W image we identify an object very close to the SN position which we consider to be the progenitor candidate. This same object is faintly detected at this position in the F303W image; see Figure~\ref{fig_progenitor}a,b. The pixel position for the SN transformed into the pre-SN F606W image is [242.83, 363.79], while the candidate position is [243.08, 363.96]; for the pre-SN F330W image, these are [348.36, 317.88] and [348.92, 318.02], respectively. The differences between the SN and the progenitor candidate positions, compared with the total estimated astrometric uncertainty, are given in Table \ref{table_progpos}. This latter uncertainty was calculated as a quadrature sum of the uncertainties in the SN and progenitor candidate positions and the RMS uncertainty in the geometric transformation.

From the results in Table \ref{table_progpos}, it can be seen that the difference between the SN position and the position of the progenitor candidate is slightly larger in right ascension for both bands than the measurement uncertainties; the agreement is much better in declination. However, the differences in position are within 3$\sigma$ of the total uncertainties. We therefore consider the candidate as the SN progenitor star and attribute the larger offset in right ascension (about 28 and 13 mas in F330W and F606W, respectively) to the complex background in the bright spiral arm of the host galaxy on which the progenitor candidate is located. We emphasize that no other point-like source exists within this 3$\sigma$ radius from the progenitor candidate position (see Figure \ref{fig_progenitor}c).

The field was observed again on 2015 October 14 as part of our Target-of-Opportunity program GO-13683 (PI S.~Van Dyk). On this occasion we obtained six dithered images of 100~s exposure with the {\sl HST\/} Wide-Field Camera 3 (WFC3) UVIS Channel ($\sim 0{\farcs}04$ pixel$^{-1}$) in F555W band (hereafter labelled as ``late-{\sl HST\/} images''). Using relative astrometry, we matched the observations of SN~2010bt before and after the explosion with this late-time drizzled {\sl HST\/} image. We did not detect any luminous point source at the SN position in the F555W image (Figure \ref{fig_progenitor}d), confirming the identification of the progenitor (see next section for further details). 

\begin{deluxetable}{cccc}
\tablenum{9}
\tablewidth{0pt}
\tablecolumns{4}
\tablecaption{Brightness\tablenotemark{a} of SN~2010bt and the Progenitor Candidate\label{table_progmag}}
\tablehead{
\colhead{} & \colhead{F330W} & \colhead{F606W} & \colhead{F555W} \\
\colhead{} & \colhead{(mag)} & \colhead{(mag)} & \colhead{(mag)}}
\startdata
Progenitor~candidate  & 24.29 (0.23)\tablenotemark{b} & 22.23 (0.02)\tablenotemark{c} & \nodata \\
SN~2010bt &   \nodata     & 22.50 (0.05)\tablenotemark{d} & $\gtrsim$24.4\tablenotemark{e} \\
\hline
\enddata
\tablenotetext{a}{Magnitude uncertainties are $1\sigma$.}
\tablenotetext{b}{Image taken with {\sl HST}+ACS/HRC on 2003 May 08.}
\tablenotetext{c}{Image taken with {\sl HST}+WFPC2 on 1994 Aug. 23.}
\tablenotetext{d}{Image taken with {\sl HST}+ACS/WFC on 2010 Oct. 09.}
\tablenotetext{e}{Image taken with {\sl HST}+WFC3/UVIS on 2015 Oct.14.}
\end{deluxetable}

\begin{deluxetable}{cccc}
\tablenum{10}
\tablewidth{0pt}
\tablecolumns{4}
\tablecaption{SN~2010bt and the Progenitor Candidate Position Comparison\label{table_progpos}}
\tablehead{
\colhead{} & \colhead{F330W} & \colhead{F606W} & \colhead{F555W} \\
\colhead{} & \colhead{($\alpha/\delta$)} & \colhead{($\alpha/\delta$)} & \colhead{($\alpha/\delta$)} }
\startdata
Total uncertainty (mas) & 10/10 & 9/8 & 10/10  \\
Diff. position SN/candidate (mas) & 28/7 & 13/8 & \nodata \\
\enddata
\tablenotetext{}{Uncertainties are $1\sigma$.}
\end{deluxetable}

%
\section{The Nature of the Progenitor Candidate}\label{natureprog}

As discussed in Section \ref{SNspec}, the early-time spectrum of SN~2010bt has a rather blue continuum, yet it also exhibits relatively strong, narrow Na\,{\sc i}~D absorption, suggesting the simultaneous presence of dust and extinction suffered by the SN. Measuring the equivalent width (EW) of the blended Na\,{\sc i}~D line at the host-galaxy redshift ($z_0 = 0.016$) from early-time optical spectrum of SN~2010bt ($\sim 1.8$ \AA), we can attempt to estimate ${E(B-V)}_{\rm tot}$. Any relationship between EW and ${E(B-V)}$ tends to break down for SNe with moderate-to-high reddening. We obtain a large dispersion in the results following the relations of different authors. Specifically, we estimate ${E(B-V)}_{\rm tot} \approx 0.3$ and $0.9$ mag (i.e., $A_V \approx 0.8$ and $2.7$ mag, respectively --- assuming the \citet{cardelli89} reddening law with updated wavelengths and $R_V = 3.1$) following \citet{turatto03}, and ${E(B-V)}_{\rm tot} \approx 1.6$ mag (i.e., $A_V \approx 5.0$ mag) using \citet{poznanski12}. Because of the large differences obtained with the EW(Na\,{\sc i}~D) vs. $E(B-V)$ relations, we decided not to use this method to estimate the extinction toward SN~2010bt. 
\citet{phillips13} have also cautioned that the EW(Na\,{\sc i}~D) vs. ${E(B-V)}$ relationship are associated with large scatter.

Thus, in order to estimate a consistent value for the extinction ($A_V$) toward SN~2010bt, we considered different methods based on comparisons of the object's SED and luminosity with those of other ``standard'' SNe~IIn. It should be noted that in fact SNe~IIn exhibit a wide diversity in both photometry and spectra (see, e.g., \citealt{kiewe12,taddia13}). We first matched simultaneously the intrinsic $(B-V)_0$, $(V-R)_0$, and $(R-I)_0$ color curves of SN~2010bt with those of other SNe~IIn (see Figure \ref{fig_color}), finding an extinction $A_V = 1.47 \pm 0.31$ mag. We also compared  the early-time optical SED of SN~2010bt (we did not used the late-time spectra, given their comparatively poor S/N) with those of the interacting SNe~1998S, 2009ip, and 2015bh at similar epochs. Spectra of the comparison SNe were first corrected for redshift and extinction, and then scaled to the distance of SN~2010bt. The average of the good matches in all cases is $A_V = 0.98 \pm 0.33$ mag. We adopt the uncertainty-weighted average value, $A_V^{\rm tot} = 1.24 \pm 0.42$ mag (i.e., $E(B-V) = 0.40 \pm 0.15$ mag), as the extinction toward SN~2010bt. 

We measured the brightness of the progenitor candidate using Dolphot, finding $m_{\rm F606W} = 22.23 \pm 0.02$ mag. Dolphot reported that the progenitor candidate has an ``object type'' flag of ``1,'' which means that the source is likely stellar. Additionally, the object had a ``sharpness'' and a ``crowding'' parameter\footnote{The ``sharpness'' parameter indicates the reliability that a detected source is indeed point-like. It is considered a ``good star" when this parameter is between $-0.3$ and $+0.3$, and a ``perfectly-fit star" when the value is $0$. The ``crowding'' parameter is defined in magnitude and describes the measured quality brightness of a star. Isolated stars have a crowding value of zero. This value increases as more stars surround the star under study, contaminating the measurement. Hence, a low value increases the certainty of the measured stars.} $\sim 0$, further indicating that the detected source is point-like and ``clean'' \citep{dolphin00}.

Dolphot also detected the candidate in F330W, although the object type is ``2,'' indicating that the source is a star too faint for a PSF determination; the other two parameters indicated that this object is an otherwise good and clean source. The magnitude measured (via aperture photometry) in this filter is $m_{\rm F330W} = 24.29 \pm 0.23$ mag. Note that the {\sl HST\/} images taken with F606W were obtained in 1994, and those using F330W in 2003.

We obtained $m_{\rm F606W} = 22.50 \pm 0.05$ mag for the SN, running Dolphot on the {\sl HST\/} images taken on 2010 (assuming that the F606W bandpass is approximately Johnson $V$, we include this measurement in Table~\ref{table_JCph}). The SN was therefore found to be slightly fainter than the progenitor candidate.

Finally, we ran Dolphot on the 2015 late-time {\sl HST\/} images of the SN field. By adding artificial stars at the SN position, we obtained conservatively an upper brightness limit for a single point source with a sigma threshold of 3.0 of $m_{\rm F555W} \gtrsim 24.4$ mag. That is, the previously identified source had disappeared, being at least 2 mag fainter than in the observations in 1994.  
The precise match between the {\sl HST\/} images taken before and after the SN explosion, the fit quality given by Dolphot, and the disappearance of the progenitor star five years after the explosion tell us that this star is likely the progenitor of SN~2010bt.
Note that the 2015 {\sl HST\/} images were obtained with WFC3, which has better throughput than WFPC2.
SN~2009ip was also observed at late time phases ($>$ 1100 d after the maximum of its brightest event 2012b), at which time the luminosity of the transient was probably still dominated by CSM interaction, and was found to be marginally fainter than its quiescent progenitor \citep{thone15,smith16a,graham17}. Unfortunately, like SN~2009ip, then, this apparent fading of SN~2010bt 15~yr after the nominally SN-like event cannot alone prove that it was a terminal explosion.

Correcting by the total extinction and distance assumed for the SN (see the beginning of this section and Section~\ref{intro}), we find that the absolute magnitude of the progenitor was $M_{\rm F606W}^0 = -12.98 \pm 0.41$ mag and $M_{\rm F330W}^0 = -11.82 \pm 0.74$ mag, while the SN was at $-12.72 \pm 0.42$ mag (F606W) in 2010 and $\gtrsim -11$ mag (F555W) in 2015. 

\begin{figure}
\figurenum{12}
\plotone{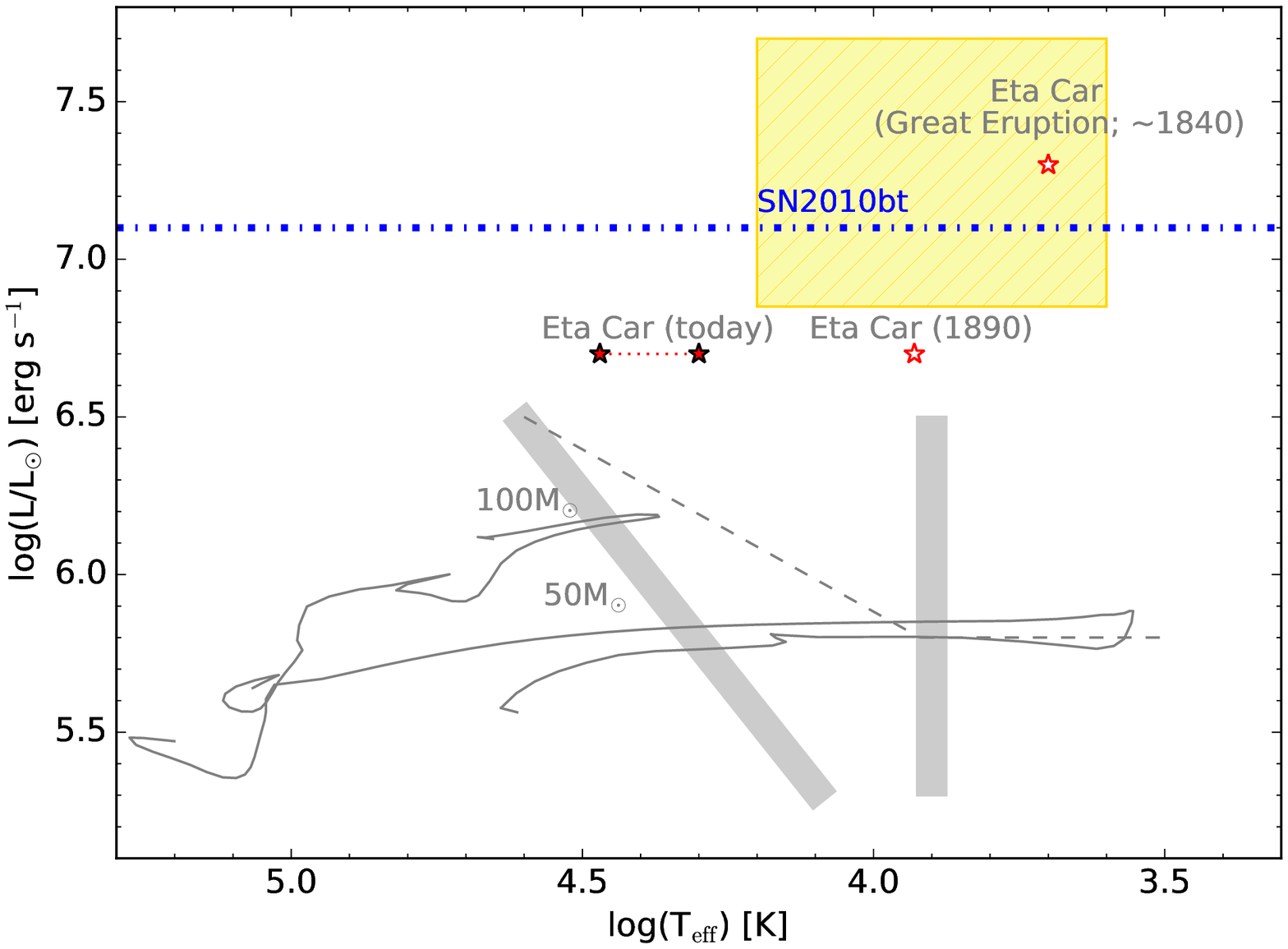}
\caption{Hertzsprung-Russell diagram showing the luminosity (dot-dashed horizontal line) of the source found at the SN~2010bt position in pre-explosion {\sl HST\/} images. For comparison, we also display the loci of the $\eta$~Carinae eruptions. The yellow (light) shaded area highlights the approximate region during the progenitor outburst phase of SN~2009ip \citep{smith10,foley11,smith14}, UGC 2773-OT \citep{smith10}, SNhunt248 \citep{kankare15}, PSN J09132750+7627410 \citep{tartaglia16b}, and SN~2015bh \citep{eliasrosa16,thone17}. Note that the color of the SN~2009ip progenitor is poorly constrained, since it was observed with only one filter. The grey (darker) shaded bands indicate the typical locations of luminous blue variables in quiescence (left, diagonal band) and during the S-Doradus-like variability (vertical band). The dashed line indicates the Humphrey-Davidson instability limit \citep{humphreys94}. We also show evolutionary tracks at 50 and 100 M$_{\sun}$ from the Cambridge STARS \citep{eldridge04} models, assuming solar metallicity.
\label{fig_hrd}}
\end{figure}

Adopting zero bolometric correction, and that $m_{\rm F606W} \approx V$, we find that the progenitor candidate has a bolometric luminosity of log($L/{\rm L}_{\sun}) \approx 7.1\pm0.2$. Such a luminosity is too large for a star in quiescence, and thus it is more likely that the progenitor was observed in eruption at that epoch.\footnote{Assuming $A_V^{\rm tot} = 0$ mag, the bolometric luminosity of SN~2010bt's progenitor is log($L/{\rm L}_{\sun}) \approx 6.8\pm0.1$. This luminosity is still more appropriate for a star in eruption.} Curiously, this value of luminosity is evocative of $\eta$~Carinae during its Great Eruption (\citealt{humphreys99,rest12}; see also Figure \ref{fig_hrd}).\\

\begin{figure}
\figurenum{13}
\epsscale{1.2}
\plotone{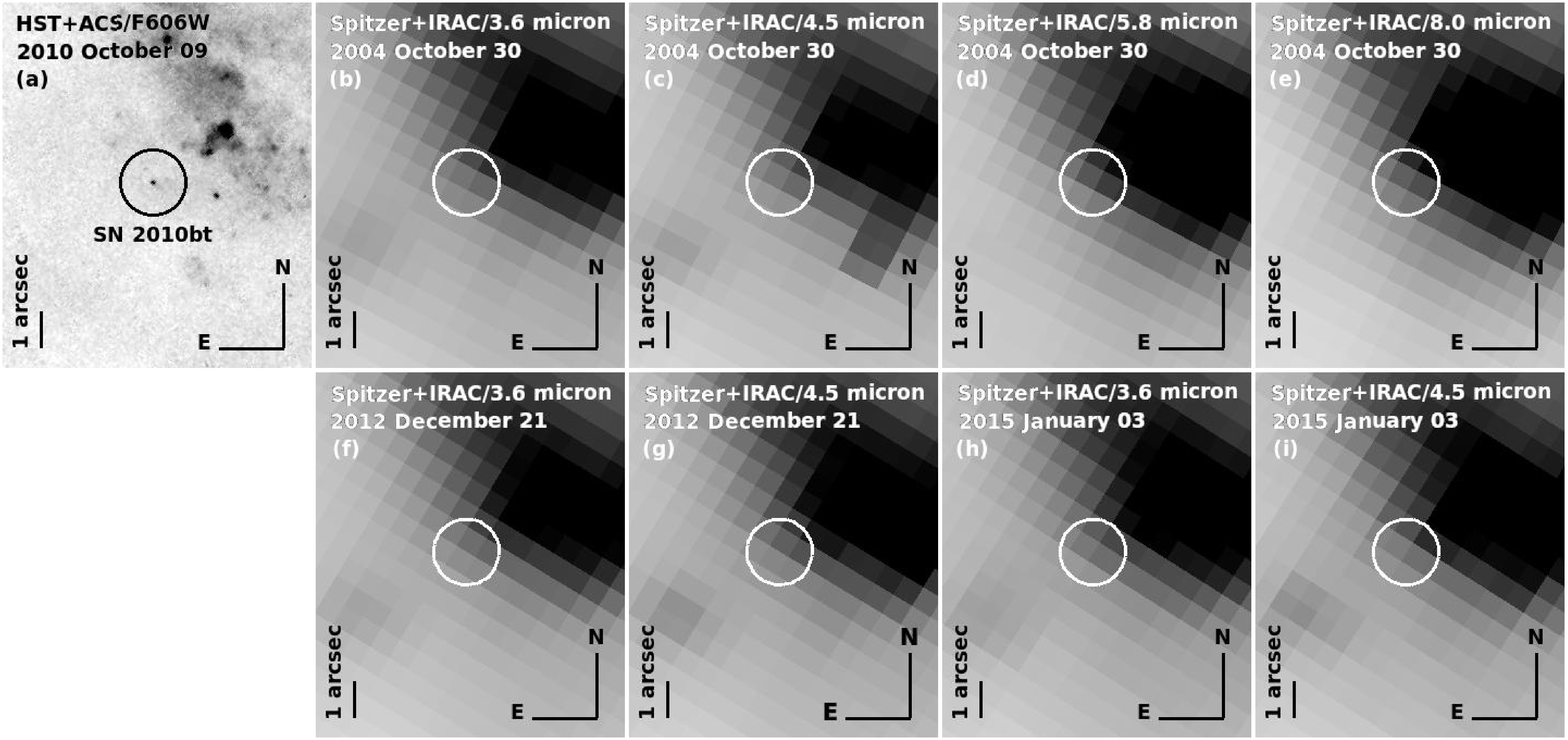}
\caption{Subsections of (a) the post-explosion {\sl HST\/} ACS/WFC image in F606W, along with (b--i) a sample of archival {\sl Spitzer} images in different channels and epochs. The position of the SN is indicated by {\it circles} with a radius of 18 pixels ($0.9''$) for the {\sl HST\/} image, and 1.5 pixels ($0.9''$) for the {\sl Spitzer} images.  \label{fig_spitzer}}
\end{figure}

It is well known that the surrounding medium of a star can affect the peak luminosity and spectra during the explosion. Indeed, SN~2010bt was classified as Type IIn owing to the presence of narrow circumstellar H emission. A possible explanation for the fact that the SN light faded so drastically in just a few months after the explosion could be the result of possible formation of dust in the ejected material, which could have obscured the SN light (or the ``remnant" if it survived the explosion). In this case, we should expect at late times a strong IR excess, as in the case of SN~1998S \citep{pozzo04} or SN~2005ip \citep[e.g.,][]{smith09,fox10}. Therefore, we analyzed several sets of IR images obtained with the InfraRed Array Camera (IRAC) onboard the {\sl Spitzer Space Telescope}, before (2004 October 30; 3.6, 4.5, 5.8, and 8 $\mu$m channels; PI J.~M.~Mazzarella; Program ID 3672) and after (2012--2015; 3.6, and 4.5 $\mu$m channels; PIs D.~B.~Sanders, O.~D.~Fox, D.~Stern; Program IDs 80089, 90031, 10098, respectively) the SN explosion. We worked with the {\it Post Basic Calibrated Data (pbcd)} provided by the {\sl Spitzer} Heritage Archive\footnote{\url{http://sha.ipac.caltech.edu/applications/Spitzer/SHA/}.}, which are already fully coadded and calibrated. Neither the progenitor candidate nor SN~2010bt was detected in any {\sl Spitzer} channel (see Figure \ref{fig_spitzer}). We therefore attempted to constrain the dust formation at the SN site by measuring the variance of the integrated flux in an area of $3 \times 3$ pixels around the progenitor/SN position using {\sc MOPEX}. Following the ``recipe'' advised by the {\sl Spitzer\/} team, we obtained the upper limits reported in Table \ref{table_spitzer}.

As one can see, there is no evidence of IR excess emission from years 2004 through 2014. 
The upper limits on the IR emission from SN~2010bt tell us that there are no detections larger than $\sim 9 \times 10^{40}$ and $\sim 7 \times 10^{40}$ erg s$^{-1}$ at $3.6$ and $4.5\, \mu$m, respectively. Thus, it is quite likely that the drop in luminosity of SN~2010bt is not caused by dust formation, but rather by the end of the ejecta interaction with the CSM.

\begin{deluxetable}{cccccccc}
\tablenum{11}
\tablewidth{0pt}
\tablecolumns{8}
\tablecaption{{\sl Spitzer\/} Flux Upper Limits at the SN~2010bt Site\label{table_spitzer}}
\tablehead{
\colhead{Date} & 
\colhead{MJD} &
\colhead{Phase\tablenotemark{a}} &
\colhead{3.6 $\mu$m} &
\colhead{4.5 $\mu$m} &
\colhead{5.8 $\mu$m} &
\colhead{8.0 $\mu$m} &
\colhead{Program ID} \\
\colhead{} & 
\colhead{} & 
\colhead{(days)} & 
\colhead{($\mu$Jy)} & 
\colhead{($\mu$Jy)} & 
\colhead{($\mu$Jy)} & 
\colhead{($\mu$Jy)} & 
\colhead{}}
\startdata
20041030 & 53308.2 & \llap{$-$}1994.9 & $< 195.5 $ & $< 196.2 $ & $< 749.2$ & $< 2560.0$ & 3672 \\
20121221 & 56282.7 & 979.5   & $< 202.6 $ & $< 209.7$ & \nodata & \nodata & 80089 \\
20121221 & 56282.9 & 979.8   & $< 169.3 $ & \nodata & \nodata & \nodata & 90031 \\
20130123 & 56315.5 & 1012.4  & $< 205.3 $ & \nodata & \nodata & \nodata & 90031 \\
20130718 & 56491.4 & 1188.3  & $< 177.3 $ & \nodata & \nodata & \nodata & 90031 \\ 
20130820 & 56524.2 & 1221.1  & $< 238.7 $ & \nodata & \nodata & \nodata & 90031 \\
20131224 & 56650.4 & 1347.3  & $< 223.1 $ & \nodata & \nodata & \nodata & 90031 \\
20140131 & 56688.6 & 1385.5  & $< 210.1 $ & \nodata & \nodata & \nodata & 90031 \\
20140727 & 56865.9 & 1562.8  & $< 193.8 $ & \nodata & \nodata & \nodata & 90031 \\
20140829 & 56898.4 & 1595.3  & $< 197.5 $ & \nodata & \nodata & \nodata & 90031 \\
20150103 & 57025.9 & 1722.8  & $< 237.6 $ & $< 177.5$ & \nodata & \nodata &  10098 \\
\enddata
\tablenotetext{a}{Phases are relative to the discovery date, MJD = 55303.1.}
\end{deluxetable}

%
\section{Was SN~2010bt a nonterminal explosion?}\label{discussion}

As discussed by many authors, the dividing line between SN impostors and real SNe is not clear, with both possibilities viable for several objects. Both families of transients share observable similarities; however, the big difference is in the progenitor star fate --- the star is destroyed in a SN explosion, but it still remains after the outburst in the case of a SN impostor. It is also hard to find a unique origin of SN~2010bt. We discuss below the observables of this transient. 
\begin{itemize}

\item Both early and late spectra show spectra dominated by strong multicomponent Balmer emission lines, indicators that define the interacting transients. The H$\alpha$ profile of SN~2010bt at early time resembles more those of the interacting transients, such as SNe~2009ip and 2015bh, than those of confirmed SNe~IIn~1996al and 1998S (see Section \ref{SNspec}). Instead at late phases, none of the comparison transients share the same H$\alpha$ profile that SN~2010bt. It could be understandable since the CSM may evolve in a different way for each object, leading to diversity among the transients. 

\item Photometrically (see Section \ref{SNph}), SN~2010bt exhibits a rapid decrease in luminosity after maximum, which is reminiscent of the SN~2009ip-like objects, and is somehow faster than other SNe. However, the luminosity at maximum of SN~2010bt is higher than SN~2009ip and most consistent with the peak luminosity of bright SNe~IIn.

\item The flattening of the late-time light curve seems slower than that expected for interacting SNe, however, it is still at the 
low end of the typical $^{56}$Ni mass range for CC-SNe. It is, however, in agreement to what was estimated for SN~2009ip-like SNe.

\item Considering the pseudobolometric luminosity of SN~2010bt, we estimate a radiated energy of at least $2 \times 10^{49}$ erg. This energy is comparable to that of SNe~IIn (e.g., \citealt{arnett96}), but also to major eruptive events of some transients, such as SN~2009ip ($\sim 3 \times 10^{49}$ erg; \citealt{margutti14}).

\item Unfortunately, except for the estimated magnitude of the progenitor star, we have not found any other data indicative of a possible pre-SN eruptive event.

\item We estimated an absolute magnitude for the progenitor star of $M_{\rm F606W}^0 \simeq -12.3$ mag (see Section \ref{natureprog}), which is similar to that derived for the SN in 2010, $-12.7$ mag (F606W). That luminosity is more like that of a star in outburst, if we compare with SN impostors, such as UGC 2773-OT \citep{smith10}, SNhunt248 \citep{kankare15}, or PSN J09132750+7627410 \citep{tartaglia16b}, which all were between $-13$ and $-14$ mag; or objects in outburst, such as SNe~2009ip (e.g. \citealt{pastorello13}), 2015bh \citep{eliasrosa16,thone17}, or $\eta$~Carinae during its Great Eruption \citep{humphreys99,rest12}. Note also that the progenitor luminosity found for the SN~IIn~2005gl \citep{galyam07,galyam09}, or some SN~2009ip-like transients \citep{smith10,foley11,smith14,eliasrosa16,pastorello18}, is around $\sim -10$ mag.

\item After several years from the discovery, the progenitor of SN~2010bt seems to have vanished, indicating that the star may had finally exploded as a SN (see Section \ref{natureprog}). This has been inferred for interacting SNe, such as SN~2005gl, but not yet for any member of the SN~2009ip-like family. As discussed by \citet{pastorello18}, SN 2009ip-like transients share similarities with SN~2005gl, supporting the possible terminal explosion scenario for all.

\end{itemize}

In short, the SN~2010bt observations have not helped us clarify whether the transient is a SN or an impostor,  with the exception of the very late-time {\sl HST\/} data that may show that the progenitor had vanished. It makes the terminal explosion scenario plausible for this event.

%
\section{Summary}\label{conclusions}

SN~2010bt was classified as an SN~IIn from an optical spectrum taken not long after the explosion. The observational campaign was interrupted after $\sim 2$ months owing to the SN's proximity to the Sun in the sky, and was again continued just a few months later. By that time, the SN had become much fainter or was nearly undetectable, which is unusual behavior for an SN in general. SN detection at late phases was obtained only through images taken with {\sl HST}, at a luminosity of $\sim -12.7$ mag in the $F606W$ band.  

Comparing {\sl HST\/} images of the host galaxy prior to the explosion and those of the SN at late times, high-precision relative astrometry allowed us to identify the likely SN progenitor star with magnitude $\sim -13$ in the $F606W$ band, which is almost certainly the luminosity of a star in eruption. At first we found that the brightness of the SN nearly five months after discovery was somewhat smaller than that estimated for the progenitor candidate. Still, no source more luminous than $-11$ mag in the $F555W$ band was found at the position of the SN five years after the SN discovery.

Overall, we have not found a unique explanation for the chain of events observed for SN~2010bt. In the following we will list our best understanding of the transient's observables.

\begin{itemize}
\item The SN~2010bt progenitor, identified in the {\sl HST\/} pre-SN images, was in outburst ($M_{\rm F606W}^0~(\sim V) \approx -13$ mag; log($L/{\rm L}_{\odot}$) $\approx 7$) at the moment of the observations in 1994. Unfortunately, no information about other possible pre-SN eruptive events has been found.

\item Some time thereafter, a powerful (terminal or nonterminal) outburst occurs, resulting in SN~2010bt. 

\item The ejecta interacted with a compact shell created during eruptions or heavy mass loss from the massive progenitor star prior to the explosion. This shock/CSM interaction led to SN~2010bt reaching a quite luminous $M_V \geq -19$ mag. 

\item A probably patchy photosphere is located in the external symmetric CSM, where the CSM-ejecta interaction is taking place, since the early-time observed H$\alpha$ profile is quite symmetric and a weak broad component is visible.

\item The CSM+ejecta recombined quickly, leading to the observed rapid decline of the SN~2010bt light curve.

\item The ejecta continued to propagate into more-distant CSM. This probably had an asymmetric geometry, since the late-time H$\alpha$ emission showed a double-peaked profile.

\item Over time, the CSM-ejecta interaction became far less strong; consequently, we saw no further trace of SN~2010bt (at least not brighter than $-11$ mag) in 2015.

\end{itemize}

In conclusion, we confirm the identification of the SN~2010bt progenitor in outburst, and favor the scenario in which SN~2010bt was a genuine SN surrounded by dense CSM with a complex geometry. \\

%
\acknowledgments 

N.E.R. thanks Avet Harutyunyan for his help, and the hospitality of the Institut de Ci\`encies de l'Espai at the Autonomous University of Barcelona's Campus, where much of this work was written. 
Support for this work was provided by NASA/{\it HST} through grants GO-11575, GO-13684, GO-14668, and AR-14295 from the Space Telescope Science Institute (STScI), which is operated by AURA, Inc., under National Aeronautics and Space Administration (NASA) contract NAS5-26555.   
S.B. and M.T. acknowledge partial financial support by the PRIN- INAF 2017 (project ``Towards the SKA and CTA era: discovery, localisation and physics of transient sources").
A.V.F. is also grateful for generous financial assistance from the Christopher R. Redlich Fund, the TABASGO Foundation, the Miller Institute for Basic Research in Science (UC Berkeley), and US NSF grant AST-1211916.
G.P. acknowledges support provided by the Millennium Institute of Astrophysics (MAS) through grant IC120009 of the Programa Iniciativa Cient\'{i}ifica Milenio del Ministerio de Econom\'{i}a, Fomento y Turismo de Chile.  
L.G. was supported in part by NSF grant AST-1311862. 
N.S. was supported in part by NSF grants AST-1312221 and AST-1515559.
This research is based in part on observations made with the NASA/ESA {\it Hubble Space Telescope}, and obtained from the Hubble Legacy Archive, which is a collaboration between STScI/NASA, the Space Telescope European Coordinating Facility (ST-ECF/ESA), and the Canadian Astronomy Data Centre (CADC/NRC/CSA); 
the {\sl Spitzer Space Telescope}, which is operated by the Jet Propulsion Laboratory, California Institute of Technology, under a contract with NASA (support was provided by NASA through an award issued by JPL/Caltech); 
the SMARTS Consortium 1.3~m telescope located at Cerro Tololo Inter-American Observatory (CTIO), Chile; and 
the New Technology Telescope at the European Southern Observatory-La Silla Observatory. 
Data in this work have been taken in the framework of the European supernova collaboration involved in ESO-NTT large program 184.D-1140 led by Stefano Benetti. Many of the spectra used here for comparison were obtained from the Padova-Asiago Supernova Archive (ASA). 
This work has made use of the NASA/IPAC Extragalactic Database (NED), which is operated by the Jet Propulsion Laboratory, California Institute of Technology, under contract with NASA; and data products from the Two Micron All Sky Survey, which is a joint project of the University of Massachusetts and the Infrared Processing and Analysis Center/California Institute of Technology, funded by
NASA and the NSF.

{\it Facilities:} \facility{Swift}, \facility{Spitzer}, \facility{ING:Kapteyn (JAG)}, \facility{CTIO:1.3m (ANDICAM)}, \facility{HST(WFPC2), ACS, WFC3}, \facility{ESO:3.6m (EFOSC2, SOFI)}, \facility{Magellan:Clay (LDSS-3)}, \facility{VLT:Yepun (HAWK-I)}

\software{IRAF \citep{tody86,tody93}, SNOoPy (http://sngroup.oapd.inaf.it/snoopy.html), SExtractor \citep{bertin96}, HOTPANTS \citep{becker15}, Dolphot \citep{dolphin00}, GELATO (https://gelato.tng.iac.es), MOPEX (http://irsa.ipac.caltech.edu/data/
SPITZER/docs/dataanalysistools/tools/mopex/)}



\bibliographystyle{aasjournal}
\bibliography{progenitors}


\end{document}